\documentclass[aps,twocolumn,floatfix,altaffilletter,superscriptaddress,preprintnumbers,
               tightenlines,showpacs,showkeys,notitlepage,nofootinbib]{revtex4-2}

\usepackage[colorlinks=true,citecolor=blue,linkcolor=blue]{hyperref}
\usepackage[normalem]{ulem}
\usepackage{amsmath,amssymb}
\usepackage{mathtools}
\usepackage{verbatim}
\usepackage{titlesec}               
\usepackage{epsfig}
\usepackage{graphicx}               
\usepackage{url}
\usepackage{color}
\usepackage{multirow}
\usepackage{placeins}
\usepackage[dvipsnames]{xcolor}
\usepackage{epstopdf}
\usepackage{siunitx}
\usepackage{fontawesome}
\usepackage{enumitem}
\usepackage{cleveref}
\usepackage{lipsum}
\usepackage{gensymb}
\usepackage{aas_macros}             
\usepackage{xspace}                 
 
\allowdisplaybreaks

\makeatletter

\renewcommand{\p@subsection}{}
\makeatother

\titleformat*{\section}{\centering\bfseries\uppercase}
\titlelabel{\thetitle\quad}
\titleformat*{\paragraph}{\bfseries}
\titlespacing*{\paragraph}{0pt}{3.25ex plus 1ex minus .2ex}{1em}

\makeatletter
\def\l@subsubsection#1#2{}
\makeatother

\newcommand{\ket}[1]{\ensuremath{| #1 \rangle}}   
\newcommand{\sprod}[2]{\ensuremath{\left\langle #1 |%
                     #2 \right\rangle}}  

\DeclarePairedDelimiter\absVal{|}{|}
\newcommand{\ev}[1]{\ensuremath{\left\langle #1 %
                     \right\rangle}} 

\newcommand{\diag}{\text{diag}}

\renewcommand{\vec}[1]{{\mathbf{#1}}}


\makeatletter
\providecommand*{\diff}%
	{\@ifnextchar^{\DIfF}{\DIfF^{}}}
\def\DIfF^#1{%
	\mathop{\mathrm{\mathstrut d}}%
		\nolimits^{#1}\gobblespace}
\def\gobblespace{%
	\futurelet\diffarg\opspace}
\def\opspace{%
	\let\DiffSpace\!%
	\ifx\diffarg(%
		\let\DiffSpace\relax
	\else
		\ifx\diffarg[%
			\let\DiffSpace\relax
		\else
  			\ifx\diffarg\{%
				\let\DiffSpace\relax
			\fi\fi\fi\DiffSpace}

\begin{document}

\title{Magnetic Moments of Astrophysical Neutrinos}

\author{Joachim Kopp}
\email{jkopp@cern.ch}
\affiliation{Theoretical Physics Department, CERN,
             1 Esplanade des Particules, 1211 Geneva 23, Switzerland}
\affiliation{PRISMA Cluster of Excellence \& Mainz Institute for
             Theoretical Physics, \\
             Johannes Gutenberg University, Staudingerweg 7, 55099
             Mainz, Germany}

\author{Toby Opferkuch}
\email{tobyopferkuch@berkeley.edu}
\affiliation{Berkeley Center for Theoretical Physics, University of California, Berkeley, CA 94720}
\affiliation{Theoretical Physics Group, Lawrence Berkeley National Laboratory, Berkeley, CA 94720}

\author{Edward Wang}
\email{edward.wang@tum.de}
\affiliation{Physik Department T70, Technische Universität München, James-Franck-Straße, 85748 Garching, Germany}

\preprint{CERN-TH-2022-213,
          MITP-22-110,
          TUM-HEP-1448/22}

\begin{abstract}
\noindent
We study the impact of neutrino magnetic moments on astrophysical neutrinos, in particular supernova neutrinos and ultra-high energy neutrinos from extragalactic sources.  We show that magnetic moment-induced conversion of Dirac neutrinos from left-handed states into unobservable right-handed singlet states can substantially change the flux and flavour composition of these neutrinos at Earth. Notably, neutrinos from a supernova's neutronisation burst, whose flux can be predicted with $\mathcal{O}(10\%)$ accuracy, offer a discovery reach to neutrino magnetic moments $\sim \text{few} \times \SI{e-13}{\mu_B}$, up to one order of magnitude below current limits. For high-energy neutrinos from distant sources, for which no robust flux prediction exists, we show how the flavour composition at Earth can be used as a handle to establish the presence of non-negligible magnetic moments, potentially down to $\text{few} \times \SI{e-17}{\mu_B}$ if the measurement can be performed on neutrinos from a single source.  In both cases, the sensitivity strongly depends on the galactic (intergalactic) magnetic field profiles along the line of sight. Therefore, while a discovery is possible down to very small values of the magnetic moment, the absence of a discovery does not imply an equally strong limit. We also comment on the dependence of our results on the right-handed neutrino mass, paying special attention to the transition from coherent deflection by a classical magnetic field to incoherent scattering on individual scattering targets. Finally, we show that a measurement of Standard Model Dirac neutrino magnetic moments, of order $\SI{e-19}{\mu_B}$, could be possible under rather optimistic, but not completely outrageous, assumptions using flavour ratios of high-energy astrophysical neutrinos.
\end{abstract}

\maketitle

\section{Introduction}
\label{sec:intro}

In 1970, Cisneros \cite{Cisneros:1970nq} first proposed the neutrino magnetic moment as an explanation for the solar neutrino problem. Since magnetic moments convert left-handed particles into right-handed ones in the presence of a magnetic field, and since right-handed neutrinos cannot be detected, such conversion leads to an apparent deficit in the measured neutrino flux. While it is now known that this deficit is due to neutrino mass mixing rather than neutrino magnetic moments, conversion between left-handed and right-handed neutrinos is still a powerful tool to constrain, or better yet, discover, neutrino magnetic moments.

In the simplest extension of the Standard Model including right-handed neutrino fields, the neutrino magnetic moment, $\mu_\nu$, is \cite{Fujikawa:1980yx}
\begin{align}
    \mu_\nu = \frac{3e G_F m_\nu}{8 \sqrt{2} \pi^2}
            \approx 3 \times 10^{-19} \mu_B \left( \frac{m_\nu}{\SI{1}{eV}} \right),
    \label{eq:munu-sm}
\end{align}
where $m_\nu$ is the neutrino mass, $e$ is the electric charge unit, $G_F$ is the Fermi constant, and $\mu_B = e / (2 m_e)$ is the Bohr magneton, with $m_e$ the electron mass.  So far, neutrino magnetic moments have proven too small to be detectable, but numerous constraints have been derived from neutrino experiments \cite{Magill:2018jla, Coloma:2017ppo} as well as from astrophysical \cite{Corsico:2014mpa, MillerBertolami:2014oki, Arceo-Diaz:2015pva, Raffelt:1999gv, Diaz:2019kim, Capozzi:2020cbu} and cosmological arguments \cite{Morgan:1981zy, Fukugita:1987uy, Elmfors:1997tt, Ayala:1999xn,  Brdar:2020quo, Carenza:2022ngg, Li:2022dkc}. While these constraints still fall several orders of magnitude short of probing magnetic moments as small as in \cref{eq:munu-sm}, they are highly relevant to theories beyond the Standard Model, where $\mu_\nu$ can take values that saturate current constraints \cite{Giunti:2014ixa, Lindner:2017uvt, Xu:2019dxe, Brdar:2020quo}. More recently, large neutrino magnetic moments have been proposed in relation to experimental anomalies, notably the XENON1T electron excess \cite{Babu:2020ivd, Miranda:2020kwy, Shoemaker:2020kji, Brdar:2020quo} which has since gone away \cite{XENONCollaboration:2022kmb}, the muon $g-2$ anomaly \cite{Babu:2021jnu}, and various $B$ physics anomalies \cite{Brdar:2020quo}.

In this paper, we will investigate the role magnetic moments play in the propagation of astrophysical neutrinos through galactic and intergalactic magnetic fields. We will consider the case of Dirac neutrinos, for which standard flavour oscillations and magnetic moment-induced spin precession can be decoupled.  Majorana neutrinos, in contrast, for which only flavour off-diagonal magnetic moments are allowed, would experience coupled spin--flavour precession, and since interstellar and intergalactic magnetic fields as well as neutrino magnetic moments are small, only negligible corrections to standard flavour oscillations are expected in this case. 

For the case of $\mathcal{O}(\SI{10}{MeV})$ neutrinos from a supernova explosion, we will show that the reduction in the detectable neutrino flux due to a magnetic moment-induced helicity flip can be large enough to be detectable for magnetic moments that are up to an order of magnitude below current limits.  This is true in particular for neutrinos from the neutronisation burst that happens early on in the explosion \cite{Janka:2006fh} because the corresponding flux can be robustly predicted with an uncertainty of only $\mathcal{O}(10\%)$ \cite{Serpico:2011ir, Wallace:2015xma, OConnor:2018sti, Kachelriess:2004ds}.  In a recent paper~\cite{Jana:2022tsa}, Jana et al.\ have studied future supernova constraints on neutrino magnetic moments as well. However, their interest was in transitions magnetic moments between active neutrino flavours, whereas we focus on magnetic moment-induced conversion between active (left-handed) and sterile (right-handed) neutrinos, relevant for instance in scenarios with Dirac neutrino mass terms. The authors of ref.~\cite{Jana:2022tsa} have considered neutrino flavour conversion in the source's magnetic field, whereas will be mainly concerned with interstellar magnetic fields. We will explain in \cref{sec:sn-general} why source magnetic fields are unimportant for us, and why interstellar magnetic fields are unimportant in ref.~\cite{Jana:2022tsa}.

In the second part of the paper, we will focus on $\mathcal{O}(\si{TeV})$ neutrinos from cosmic ray accelerators, which may experience a similar helicity flip during propagation. However, as there is no robust prediction for their initial flux, the resulting flux reduction is undetectable. Nevertheless, if magnetic moments are flavour non-universal, the disappearance of only some neutrino flavours may be detectable in measurements of neutrino flavour ratios.

The structure of this paper is as follows: in \cref{sec:mm} we introduce the notation and present approximate analytic results for the conversion probability from left-handed to right-handed neutrinos in a magnetic field. In \cref{sec:B-fields} we describe the properties of interstellar and intergalactic magnetic fields pertinent to our study. In \cref{sec:sn} we then discuss the neutrino flux from supernovae, in particular during the neutronisation phase, and apply the results from \cref{sec:mm,sec:B-fields} to this flux.  This allows us to estimate the discovery reach  for neutrino magnetic moments in DUNE and Hyper-Kamiokande. We then move on in \cref{sec:uhe} to ultra-high energy neutrinos detectable at neutrino telescopes, where in particular flavour non-universal magnetic moments can dramatically affect the observed flavour ratios. We finally broaden our discussion in \cref{sec:discussion} to comment on the prospects of detecting neutrino magnetic moments as small as in the Standard Model, and to elaborate on how the mass of the right-handed neutrinos affects our results. In this context, we discuss the transition from coherent deflection by a classical magnetic field to hard scattering on individual electrons/nuclei in the interstellar medium. We conclude in \cref{sec:conclusions}.

All numerical codes used in this paper, as well as all plots, can be found on GitHub \cite{GitHubCode}.

\section{Neutrino Magnetic Moments}
\label{sec:mm}

Neutrino magnetic moments are described by the operator
\begin{align}
    \mathcal{L} \supset \frac{\hat{\mu}_\nu^{\alpha\beta}}{2} F_{\mu \nu} \bar{\nu}_L^\alpha \sigma^{\mu \nu} N_R^\beta + \text{h.c.},
    \label{eq:Lagrangian}
\end{align}
where $\nu_L^\alpha$ is the left-handed neutrino field of flavour $\alpha$, $N_R^\beta$ are right-handed neutrino fields, and $F_{\mu \nu}$ is the electromagnetic field strength tensor. In the case of Majorana neutrinos, $N_R$ can be the charge conjugate of a left-handed neutrino field, $N_R^\beta = (\nu^\beta)^c$. In this case, the CPT-symmetry mandates that the magnetic moments of particles and antiparticles have opposite signs, meaning that the magnetic moment matrix $\hat{\mu}$ is antisymmetric and, in particular, that its diagonal components vanish. While the numerical results we present in \cref{sec:sn,sec:uhe} will be applicable only to Dirac neutrinos (as competitive limits arise only in this scenario), we will keep the discussion in this section general and include also the Majorana case. As usual, the neutrino flavour eigenstate fields $\nu_L^\alpha$ are related to their mass eigenstate counterparts, $\nu^i$, according to
\begin{align}
    \nu^\alpha = U_{\alpha i} \nu^i \,,
\end{align}
where
\begin{align}
    U = \begin{pmatrix}
          1 &         &        \\
            &  c_{23} & s_{23} \\
            & -s_{23} & c_{23}
        \end{pmatrix}\!\!\!
        \begin{pmatrix}
          c_{13}              &   & s_{13} e^{i\delta} \\
                              & 1 &                    \\
          -s_{13} e^{i\delta} &   & c_{13}
        \end{pmatrix}\!\!\!
        \begin{pmatrix}
          c_{12} & s_{12} &   \\
         -s_{12} & c_{12} &   \\
                 &        & 1
        \end{pmatrix} \notag
\end{align}
is the leptonic mixing matrix. It depends on the neutrino mixing angles $\theta_{ij}$ through $s_{ij} \equiv \sin\theta_{ij}$, $c_{ij} \equiv \cos\theta_{ij}$, and on the CP-violating phase $\delta$.

The flavour evolution of a neutrino is governed by the Schr\"odinger equation
\begin{align}
    i \frac{\text{d}}{\text{d}t} \psi(t) = \hat{H} \psi(t) \,,
    \label{eq:schroedinger}
\end{align}
where $\psi$ is a unit vector in flavour space whose components describe the admixture of each flavour to the neutrino state. Thus, for the case of three right-handed states, $\psi = (\nu^e, \nu^\mu, \nu^\tau, N_R^e, N_R^\mu, N_R^\tau)$. In this case, a pure $\nu_\mu$ would be described by the vector $\psi = (0,1,0,0,0,0)$.  If neutrinos travel along the $z$-axis through a magnetic field $\vec{B}$, the Hamiltonian in the flavour basis can be written in block-diagonal form as\footnote{This can be most easily seen by direct evaluation of \cref{eq:Lagrangian}, choosing a particular representation for the spinors and $\gamma$ matrices.}
\begin{align}
    \hat{H} &= \frac{1}{2p}
               \begin{pmatrix}
                   U &   \\
                     & U
               \end{pmatrix}
               \begin{pmatrix}
                   \hat{M}_\nu^2 & \\
                                 & \hat{M}_N^2
               \end{pmatrix}
               \begin{pmatrix}
                   U^\dag & \\
                          & U^\dag
               \end{pmatrix}
                                    \notag\\[0.2cm]
            &\hspace{2cm}
               +
               \frac{1}{2}
               \begin{pmatrix}
                   0 & B_\perp e^{i\phi} \hat{\mu}_\nu \\
                   B_\perp e^{-i\phi}\hat{\mu}_\nu^\dagger & 0
               \end{pmatrix} \,,
    \label{eq:H}
\end{align}
with the neutrino momentum $p$, the diagonal mass matrices
\begin{align}
    \hat{M}_\nu &= \diag( m_{\nu_1}, m_{\nu_2}, m_{\nu_3} ) \,,  \label{eq:M-nu} \\
    \hat{M}_N   &= \diag( m_{N1},    m_{N2},    m_{N3} )    \,,  \label{eq:M-N}
\end{align}
and the general magnetic moment matrix (with the individual entries in this matrix labelled using the notation defined in \cref{eq:Lagrangian})
\begin{align}
    \hat{\mu}_\nu = \begin{pmatrix}
                        \mu_\nu^{ee}     & \mu_\nu^{e\mu}    & \mu_\nu^{e\tau}    \\
                        \mu_\nu^{\mu e}  & \mu_\nu^{\mu\mu}  & \mu_\nu^{\mu\tau}  \\
                        \mu_\nu^{\tau e} & \mu_\nu^{\tau\mu} & \mu_\nu^{\tau\tau}
                    \end{pmatrix} \,.
    \label{eq:mu}
\end{align}
We have written the components of the magnetic field perpendicular to the neutrino's direction of travel in terms of $B_\perp \equiv \sqrt{B_x^2 + B_y^2}$ and $\phi \equiv \arctan(B_x/B_y)$.\footnote{The longitudinal component of $\vec{B}$ is unimportant here as can be seen from the transformation of the magnetic field to the neutrino's rest frame.  For ultra-relativistic neutrinos, $B_\perp$ in the rest frame is enhanced by a relativistic $\gamma$-factor, while the longitudinal component, $B_\|$, is not.} If the magnetic field direction is constant along the line of sight, the phase $\phi$ is unphysical. For spatially varying magnetic fields, however, it is important as we will discuss below.

As astrophysical neutrinos propagate as mass eigenstates,\footnote{This can be understood by considering that the oscillation lengths corresponding to standard flavour oscillations are much shorter than the typical distances that astrophysical neutrinos travel. Neighbouring oscillation maxima are therefore extremely close in energy and cannot be resolved by the detector. In the mass basis, the off-diagonal elements of the density matrix describing the observed neutrino flux therefore average to zero, implying an incoherent statistical mixture of mass eigenstates.} it is useful to rotate $\hat{H}$ to the mass basis
\begin{align}
    \hat{H} \!=\! \frac{1}{2p}
              \begin{pmatrix}
                  \hat{M}_\nu^2 & \\
                                & \hat{M}_N^2
              \end{pmatrix}
             \!\! + \!
              \frac{1}{2}
              \begin{pmatrix}
                  0 & B_\perp e^{i\phi} \hat{\tilde\mu}_\nu \\
                  B_\perp e^{-i\phi}\hat{\tilde\mu}_\nu^\dagger & 0
              \end{pmatrix} \,,
    \label{eq:H2}
\end{align}
where $\hat{\tilde\mu}_\nu \equiv U^\dag \hat\mu_\nu U$.

While numerically solving \cref{eq:schroedinger} is straightforward, deriving an approximate analytic solution provides additional insight. Utilising the two-flavour approximation—considering only one left-handed neutrino, $\nu_L$, and one right-handed neutrino, $N_R$, with a transition magnetic moment $\mu_\nu$—and under the assumption of a uniform magnetic field, we ascertain the conversion probability to be
\begin{multline}
    P_{\nu_L \to N_R}(\mu_\nu; t) = \frac{4 p^2 \mu_\nu^2 B_\perp^2}{(\Delta m^2_{N\nu})^2 + 4 p^2 \mu_\nu^2 B_\perp^2} \\
    \sin^2 \bigg( \frac{\sqrt{(\Delta m^2_{N\nu})^2 + 4 p^2 \mu_\nu^2 B_\perp^2}}{4 p} t \bigg) \,,
    \label{eq:P-LtoR-2f}
\end{multline}
with $\Delta m^2_{N\nu} = m_N^2 - m_\nu^2$.  Before discussing the conditions under which the two-flavour approximation is useful, let us first interpret \cref{eq:P-LtoR-2f}.  Both the oscillation amplitude and the oscillation phase are governed by the interplay of the frequency $\Delta m_{N\nu}^2/(2p)$ and the magnetic moment interaction, $\mu_\nu B_\perp$.  In the case of $m_N = m_\nu$ (Dirac neutrinos), the oscillation amplitude becomes maximal, and the conversion probability reduces to 
\begin{align}
    P_{\nu_L \to N_R}(\mu_\nu; t) \xrightarrow{\Delta m_{N\nu}^2 \to\, 0} \sin^2 \frac{\mu_\nu B_\perp \, t}{2} \,.
    \label{eq:P-LtoR-Dirac}
\end{align}
In the opposite limit, $\Delta m_{N\nu}^2/(2p) \gg \mu_\nu B_\perp$, left-handed and right-handed neutrinos oscillate into one another with a frequency determined by their masses, $\Delta m_{N\nu}^2/(4p)$. The oscillation amplitude, however, becomes very small in this limit.

These observations already hint at one of the reasons why the two-flavour approximation is useful: in the case of Dirac neutrinos, conversion between the left-handed and right-handed components of the same Dirac spinor have maximal amplitude, while transitions between states of different mass are strongly suppressed. If the magnetic moment matrix $\hat{\tilde\mu}$ in \cref{eq:H2} is diagonal in the mass basis, standard flavour oscillations and magnetic moment-induced conversion between left-handed and right-handed states decouple completely. But even for non-zero flavour off-diagonal magnetic moments, the mixing between different Dirac neutrino mass eigenstates during propagation is negligible given the weakness of large-scale magnetic fields in the Universe and the strong constraints on neutrino magnetic moments. In other words, Dirac neutrinos that form an incoherent ensemble of mass eigenstates will predominantly experience mixing only between the two Weyl components of each mass eigenstate and mixing between different mass eigenstates is negligible. (We have explicitly verified this numerically.) The description in terms of an incoherent ensemble of mass eigenstates is always appropriate for astrophysical neutrinos. First because the size of the neutrino production region is typically similar in size to, or larger than, the oscillation length, and second because any coherence between mass eigenstates will quickly be lost during propagation as wave packets separate.\footnote{Note that wave packet separation between the two components of a Dirac neutrino due to energy splitting in the magnetic field is not an issue. The velocity difference between two states of mass $m$ and energies $E$ and $E + \Delta E$ is $\Delta v = m^2 \Delta E / (E^2 \sqrt{E^2 - m^2})$, and the distance by which the energy eigenstate wave packets get separated after travelling a distance $L$ is $\Delta x = L \, \Delta v$. This difference is much smaller than even the most conservative estimate for the neutrino wave packet size ($\sigma_x \sim \SI{1}{\angstrom}$) for both supernova neutrinos ($E \sim \SI{30}{MeV}$, $L \sim \SI{10}{kpc}$) and ultra-high-energy neutrinos ($E \gtrsim \SI{1}{TeV}$, $L \sim \si{Gpc}$).}

The above discussion also illustrates in more detail why the methods discussed in this paper are suitable only for constraining magnetic moments of Dirac neutrino  (and even for those only the diagonal components of the magnetic moment matrix in the mass basis), not those of Majorana neutrinos.  For astrophysical Majorana neutrinos propagating as an incoherent ensemble of mass eigenstates, the small magnetic moment-induced changes in the mixing between these mass eigenstates will be negligible compared to standard three-flavour mixing.

The other approximation entering \cref{eq:P-LtoR-2f}, namely that of constant $B$-field, is typically not applicable to astrophysical environments (see \cref{sec:B-fields} below for a detailed discussion). It is still useful for obtaining order-of-magnitude estimates, though. To go beyond these, \cref{eq:schroedinger} needs to be integrated numerically, which is what we will do in \cref{sec:sn,sec:uhe} using the Python packages \texttt{QuTiP}~v4.7.0 \cite{Johansson:2012, Johansson:2013}, \texttt{SciPy}~1.7.1~\cite{Virtanen:2020}, and \texttt{NumPy}~v1.19.5 \cite{Harris:2020}. (For propagation in turbulent magnetic fields, we also outline in \cref{sec:Posc-approx} a numerical shortcut based on the stochastic nature of the problem.) 
Analytical insights can still be gained in some special cases. One such useful case arises if only the magnitude of the transverse magnetic field changes, but not its direction The oscillation probability in the limit $\mu_\nu B_\perp \gg \Delta m_{N\nu}^2 / (2p)$ is
    \begin{align}
        P_{\nu_L \to N_R}(\mu_\nu; t)|_{\phi = \text{const}}
            \simeq \sin^2 \bigg(\! \frac{\mu_\nu}{2}\!  \int_0^L \!\!\! dx \, B_\perp(x) \!\bigg) \,,
        \label{eq:P-LtoR-varying-B-1}
    \end{align}
which can be understood as a generalisation of \cref{eq:P-LtoR-Dirac}. Of course, in reality, both the magnitude and the direction of the field are expected to vary. For further theoretical considerations and phenomenological implications of the neutrino magnetic moment, see refs. \cite{Akhmedov:1993sh, Akhmedov:1997yv, Broggini:2012df, Kurashvili:2020nwb, Popov:2019nkr} and references therein.

\section{Interstellar and Intergalactic Magnetic Fields}
\label{sec:B-fields}

The neutrino magnetic moment will interact with the magnetic fields it encounters as it propagates. For supernova burst neutrinos, which are detectable in large numbers only from within the Milky Way, we need to model the galactic magnetic field, whereas for high-energy astrophysical neutrinos, which mostly come from outside our galaxy, extragalactic magnetic fields are more important due to their huge travel distance.

\subsection{The Milky Way's Magnetic Field}
\label{sec:B-field-MW}

The Milky Way's magnetic field has both a large-scale component that is coherent on length scales of order \si{kpc}, and a small-scale component that is turbulent \cite{Beck:2007, Beck:2008ty, Haverkon:2014, Klein:2015, Ferriere:2015}. The large-scale magnetic field tends to follow the spiral arms of our galaxy. To describe this, one defines the pitch
\begin{align}
    p = \arctan\bigg( \frac{B_r}{B_\phi} \bigg) \,,
\end{align}
where $B_r$ is the radial component of the magnetic field in the galactic plane, and $B_\phi$ is its azimuthal component. The shape of the Milky Way's spiral arms is well-described by a logarithmic spiral, that is, by a parametric curve of the form $r \propto e^{k \phi}$, where $r$ and $\phi$ are polar coordinates in the galactic plane and $k$ is a constant. Such a curve is characterised by a constant pitch $p = \arctan k$. For the Milky Way, estimates for $p$ vary between $-5^\circ$ and $-30^\circ$, and the strength of the coherent magnetic field is 1.5--$\SI{2}{\mu G}$ in the solar neighbourhood and increases towards the Galactic Centre \cite{Haverkon:2014}.

\begin{figure}
    \centering
    \includegraphics[width=\columnwidth]{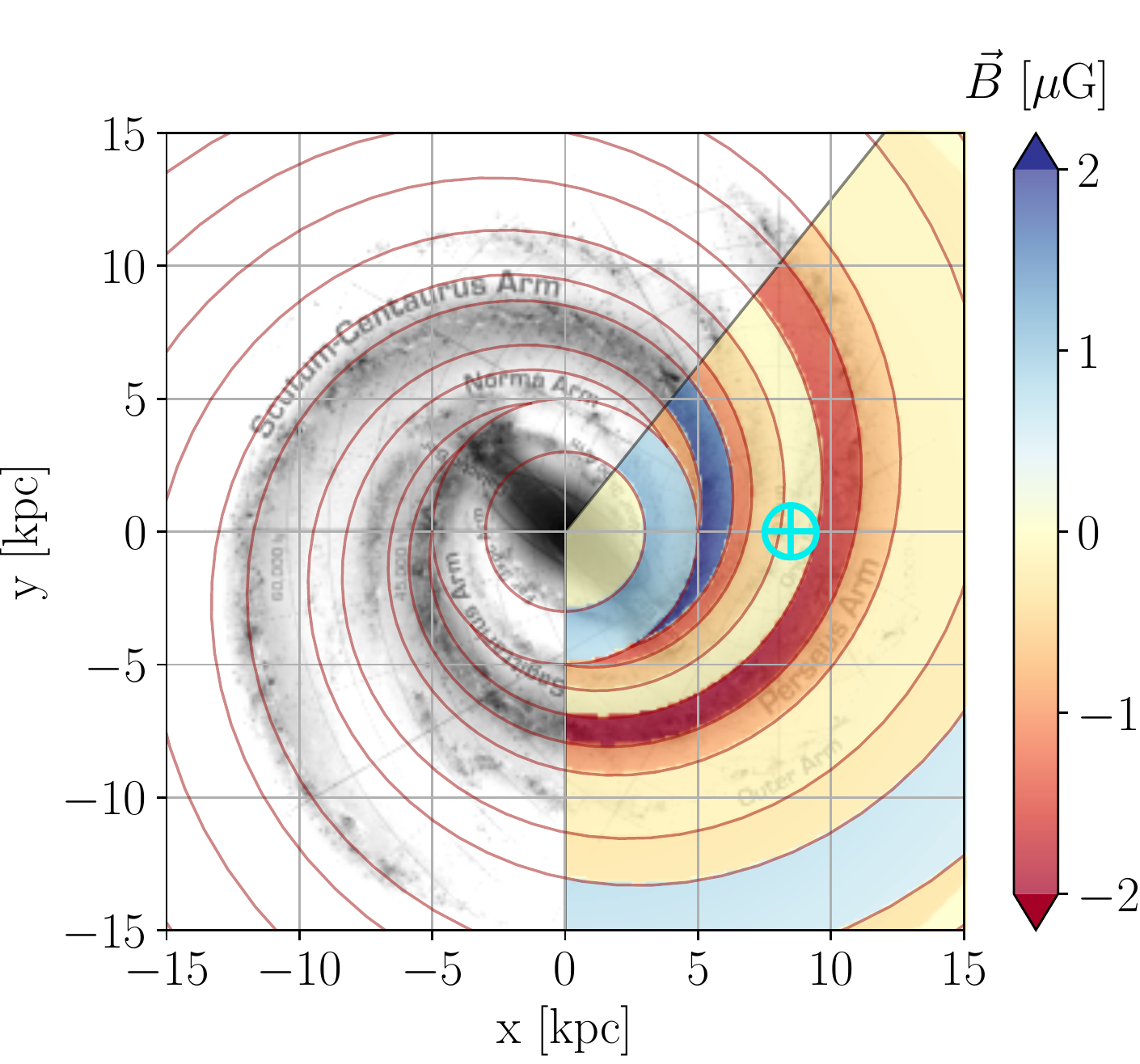}
    \caption{The coherent component of the Milky Way's magnetic field according to the model from ref.~\cite{Brown:2007qv} which we adopt here. We indicate the log-spirals delineating the Milky Way's spiral arms, with the shading indicating the magnetic field strength. Shades of blue indicate magnetic fields pointing in the counterclockwise direction, while shades of orange/red indicate fields pointing clockwise. The location of our solar system is indicated by the cyan $\oplus$ symbol. Note that this figure is rotated by 90 degrees clockwise compared to fig.~4 of ref.~\cite{Brown:2007qv}.}
    \label{fig:B-field}
\end{figure}

As a ``fiducial'' model for the large-scale galactic magnetic fields, we will use the one from ref.~\cite{Brown:2007qv} (see refs.~\cite{Sun:2007mx, VanEck:2010ka, Akahori:2017lhe} for alternative models), which divides the Milky Way into seven regions delimited by eight logarithmic spiral curves, as shown in \cref{fig:B-field}. The starting point of each spiral is set at a galactic radius of \SI{5}{kpc}. The central \SI{3}{kpc} are assumed to be free of large-scale fields, and between galactic radii of \SI{3}{kpc} and \SI{5}{kpc} the field is proportional to $1/r$ and oriented azimuthally. In the seven spiral-shaped regions, the field strength follows a $1/r$ law as well, but with a different coefficient in each region. The field is oriented tangentially to the logarithmic spiral curves. To account for uncertainties in the large-scale galactic magnetic field, we will also consider scenarios in which the coefficient of the $1/r$ law in each domain of the model is chosen randomly according to a Gaussian of width $\SI{1}{\mu G}$, centred at zero magnetic field.

The turbulent component of the Milky Way's magnetic field can be characterised by its spectral density function, $P(k)$, which describes how the energy stored in the field is distributed across eddies at different scales $k$. It is defined implicitly in terms of the two-point correlation function as
\begin{align}
    \ev{B(\vec{X} + \Delta X) B(\vec{X})}
      = \int\!\text{d}^3k \, e^{i \vec{k} \vec{X}} P(k) \,.
\end{align}
Note that we assume here large-scale isotropy and homogeneity, so that $P(k)$ depends only on $k \equiv |\vec{k}|$, but not on any angular variables. It is easy to see that $P(k) = |\tilde{B}(k)|^2$, where $\tilde{B}(k)$ is the three-dimensional Fourier transform of $B(\vec{X})$.  It is reasonable to assume that, similar to innumerable other turbulent phenomena in astrophysics and elsewhere, the turbulent magnetic fields in the Milky Way obey Kolmogorov's theory of turbulence \cite{Kolmogorov:1941}, which stipulates that the energy carried by eddies at scale $k$ is proportional to $k^{-5/3}$.  This implies that $P(k)$ scales as
\begin{align}
    P(k) \propto k^{-11/3} \,.
    \label{eq:P-k}
\end{align}
The spectral density function follows this scaling at scales smaller than some cutoff, which is between a few parsec to \SI{100}{pc} \cite{Haverkon:2014}. The magnitude of turbulent galactic magnetic fields is of order few $\si{\mu G}$ \cite{Haverkon:2014}.

For our fiducial model, we choose the cutoff of Kolmogorov scaling at \SI{10}{pc}, and the magnitude of the field at $\SI{2}{\mu G}$. We divide the frequency domain into 1000 equidistant bins and choose the power in each of them according to a Gaussian distribution with width given by \cref{eq:P-k} and centred at zero. The phase in each bin is chosen randomly between zero and $2\pi$. We then apply a Fast Fourier Transform to obtain the (discretised) magnetic field configuration along the line of sight. We normalise this field configuration such that the root-mean-square magnetic field strength is $\SI{2}{\mu G}$. This field configuration is then added to the large-scale magnetic field whose modelling has been discussed above, and fed into the evolution equation, \cref{eq:H2}. To account for uncertainties, we consider a large number of field configurations randomly generated with this method, and we also vary the cutoff between \SI{1}{pc} and \SI{100}{pc}, and the root-mean-square field strength according to a Gaussian distribution with a width of $\SI{5}{\mu G}$.

Much of our knowledge about galactic magnetic field is derived from measurements of Faraday rotation \cite{Han:2017} of radio waves. When linearly polarised electromagnetic waves cross a magnetised medium, the polarisation angle changes proportional to $\lambda^2 \int \! ds \, n_e(s) B_\|(s)$, where $\lambda$ is the photon wavelength, $n_e(s)$ is the electron number density, $B_\|(s)$ is the magnetic field component parallel to the photon momentum, and the integral runs along the line of sight. The integral also defines the so-called ``rotation measure'' or RM (up to an $\mathcal{O}(1)$ constant). Suitable sources for RM studies include pulsars as well as extragalactic sources such as quasars or fast radio bursts.  As the polarisation angle of the source is not known, the $\lambda$-dependence of Faraday rotation is used to extract the RM from observations of the photon polarisation at different wavelengths.

\subsection{Extragalactic Magnetic Fields}
\label{sec:B-field-extragalactic}

Relatively little is known about magnetic fields in between galaxies \cite{Klein:2015, Han:2017}. Inside galaxy clusters, observations of diffuse radio emission and of the polarised photons from radio sources seen through a cluster provide clues and point towards field strengths of order \si{\mu G} (with $\mathcal{O}(1)$ uncertainties) \cite{Han:2017}. There are also more indirect ways of constraining intracluster fields using the (non)observation of x-rays from inverse Compton scattering induced by the high-energy electron population needed to explain the observed synchrotron emission \cite{Bartels:2015kpa}. The lower limits on $|\vec{B}|$ derived in this way are roughly consistent with the values derived from  diffuse radio emission and from the radiation measure of radio sources seen through a cluster. Observations also indicate that intracluster fields should be irregular and feature abundant small-scale structures.

Evidence for magnetic fields outside of galaxy clusters is scarce \cite{Han:2017} and partly contradictory. On the one hand, synchrotron emission from a region in between galaxy clusters suggests a magnetic field of order 0.2--$\SI{0.6}{\mu G}$ \cite{Kim:1989, Brown:2011, Kronberg:2007wa}, which would be only an order of magnitude lower than galactic magnetic fields. On the other hand, the dependence of sources' rotation measures on redshift suggests intergalactic magnetic fields cannot be larger than $\mathcal{O}(\si{nG})$~\cite{Pshirkov:2015tua}. Weak \emph{lower} limits on extragalactic magnetic fields of order $|\vec{B}| \gtrsim$ \SIrange{e-16}{e-13}{G} can also be derived from gamma-ray observations of blazars, see ref.~\cite{Han:2017} and references therein. High-energy primary photons from such objects should undergo secondary interactions in the intergalactic medium, leading to electromagnetic cascades (elongated over astrophysical distance scales) and thus to secondary gamma-ray emission at lower energies. The non-observation of this lower-energy emission can be understood by arguing that secondary electrons and positrons are deflected away from the line of sight by magnetic fields.

A possible game-changer in the study of extragalactic magnetic fields are fast radio bursts (FRBs) \cite{Akahori:2016ami, Vazza:2018fbb, Bhandari:2021}. They are observable out to very large redshifts, and a large sample has recently become available from the CHIME/FRB collaboration \cite{CHIMEFRB:2021srp}. Individual FRBs have already been used to derive constraints on the order of $|\vec{B}| < \text{few} \times \SI{10}{nG}$ \cite{Ravi:2016kfj, Bannister:2019iju}.

We define our fiducial model of intergalactic magnetic fields in the same way as the one for turbulent galactic magnetic fields. However, we divide the line of sight into three regions, corresponding to propagation through the galaxy cluster in which the source is located, the intercluster space, and our local galaxy cluster. The travel distance within each cluster is taken to be \SI{10}{Mpc}, while the total distance is varied between 0.1 and \SI{5}{Gpc}. We set the outer cutoff of the Kolmogorov spectrum to \SI{1}{Mpc} within galaxy clusters, and to \SI{10}{Mpc} outside. The root-mean-square strength of the intracluster fields is drawn from a normal distribution with a width of $\SI{1}{\mu G}$, while for the intercluster fields we use a width of \SI{5}{nG}.

\section{Supernova Neutrinos}
\label{sec:sn}

\subsection{General Considerations}
\label{sec:sn-general}

In this section we demonstrate how neutrinos from the neutronisation (or deleptonisation) burst phase of a galactic core-collapse supernova can be used to probe neutrino magnetic moments. The neutronisation burst proceeds, as the name suggests, largely through the process $e^- + p \to \nu_e + n$. The neutrinos are initially trapped, but leak out of the supernova core in the first $\sim \SI{20}{ms}$ after core bounce. During this phase, the luminosity reaches a spectacular \SI{e53}{erg/sec}.  In contrast to neutrinos from the subsequent accretion and cooling phases, the emission during the neutronisation burst is dominated completely by the production of $\nu_e$-flavoured neutrinos \cite{Janka:2006fh, Wallace:2015xma, OConnor:2018sti}.

The crucial point for us is that the $\nu_e$ flux emitted during the neutronisation burst can be fairly well predicted (at the 10\% level) as the properties of a stellar core at the point where it reaches the Chandrasekhar threshold and collapses are similar in all core-collapse supernovae \cite{Kachelriess:2004ds}. This means that a moderate deficit of neutrinos, caused by their conversion into invisible right-handed states due to a magnetic moment, could be observable. Whether the $\nu_L$ deficit is large enough to be detected depends on the interplay of the mass squared difference $\Delta m_{N\nu}^2$ and the magnetic potential $\mu_\nu B_\perp$, in combination with the neutrinos' typical travel distance, $L \sim \SI{10}{kpc}$, and their $\mathcal{O}(\SI{10}{MeV})$ energy.

As stated in \cref{sec:mm} we will focus on the case of pure Dirac neutrinos where $\Delta m_{N\nu}^2=0$ for two main reasons: ($i$) This  trivially satisfies the requirement that the amplitude of the magnetic moment induced oscillations are large $\Delta m_{N\nu}^2 / (\mu_\nu^2 B_\perp^2) \lesssim 1$, c.f. \cref{eq:P-LtoR-2f}. ($ii$) It removes the dependence of the oscillation frequency on the momentum of the neutrino, see \cref{eq:P-LtoR-Dirac}. Before turning to simulations we can estimate the range of $\mu_\nu$ where observable effects occur. Given the amplitude is maximised, we must simply study the oscillation length, which from \cref{eq:P-LtoR-Dirac} is $L_\text{conv} = 2 \pi / (\mu_\nu B_\perp)$. This implies that only a certain range of $\mu_\nu$ can be probed for fixed values of $L$ and $B_\perp$. The lower limit arises from the requirement that the oscillation phase is not vanishingly small, while an upper limit occurs from requiring $L \lesssim L_\text{conv}$, yielding $\mu_\nu \in [\SI{3}{}, \SI{70}{}]\times10^{-14}\mu_B$. This range assumes a constant magnetic field with $B_\perp = \mu\text{G}$ and $L = \SI{10}{kpc}$.

Note that, throughout our discussion, we neglect $\nu_L \leftrightarrow N_R$ conversion inside the exploding star. At first, this might seem surprising, given that magnetic fields outside a supernova core can reach \SI{e10}{G}, so according to \cref{eq:P-LtoR-Dirac} we could expect sensitivity to $\mu_\nu \lesssim \SI{e-14}{\mu_B}$ from conversion in the supernova core alone. However, the large matter density in and around the supernova core suppresses $\nu_L \leftrightarrow N_R$ oscillations via the Mikheyev--Smirnov--Wolfenstein (MSW) effect \cite{Wolfenstein:1977ue, Mikheyev:1986gs, Mikheyev:1986wj}. The MSW potential generated by coherent forward scattering of neutrinos on ambient nucleons and electrons, $|V_\text{MSW}| \sim G_F n_n$, where $n_n$ is the ambient neutron density, affects $\nu_L$ but not $N_R$. It therefore produces a large contribution to the upper left block of the Hamiltonian in \cref{eq:H,eq:H2}, thus suppressing the effective mixing angle between $\nu_L$ and $N_R$. In this case resonant transitions are nevertheless possible \cite{Lim:1987tk,Akhmedov:1987nc,Akhmedov:1988uk}, particularly if the MSW potential for electron neutrinos $V_\text{MSW} \sim G_F (n_e - n_n/2)$ is small. However, in our simulations we do not find regions inside the SN where these resonant transitions are relevant.\footnote{Note however, that the MSW potential neutrinos encounter while propagating through the dilute interstellar medium is negligible, even compared to the neutrinos' tiny magnetic interaction.} Note that this is different from the situation discussed for instance in refs.~\cite{Lim:1987tk, Akhmedov:1996ec, Totani:1996wf, Nunokawa:1996gp, Ando:2003pj, Akhmedov:2003fu, Yoshida:2009ec}, as well as the recent ref.~\cite{Jana:2022tsa}, where magnetic moment-induced $\nu_L \leftrightarrow N_R$ conversions are a dominant effect. This is because the authors of \cite{Jana:2022tsa} are interested in transition magnetic moment between different active Majorana neutrino flavours. The MSW potential is approximately the same for all active neutrino species, so its effect cancels out to high accuracy. Lastly, we also ignore any possible effects arising from the rotation of the magnetic field \cite{Vidal:1990fr, Smirnov:1991ia, Akhmedov:1991vj}. A recent paper \cite{Jana:2023ufy} studies these effects for SN with extremely strong magnetic fields at the surface of the Iron core. Resonant conversion could occur, but requires \SI{e+12}{G} magnetic fields which are moreover strongly twisted, with order-one rotation of the field over kilometer distance scales. It is currently unclear whether such twists in the SN magnetic field exist at the time of core bounce. Nevertheless we emphasise that resonances from this effect can lead to additional time dependence in the observed events, beyond what is presented below.

\subsection{Expected Change in Event Rates}
\label{sec:sn-rate}

We model the initial neutrino flux emitted from the supernova core according to the results of ref.~\cite{Hudepohl:2009tyy}, which correspond to an electron-capture supernova with a $\SI{8.8}{M_\odot}$ progenitor star.  As we consider Dirac neutrinos, for which we have seen in \cref{sec:mm} that the $\nu_L \to N_R$ conversion probability is independent of energy, the neutrino spectrum is not relevant to the following discussion -- only the flux as a function of time matters.

To account for neutrino flavour transitions on the way out of the supernova core, we assume simple adiabatic conversion \cite{Dighe:1999bi, Scholberg:2017czd}. The flux of flavour $\nu_\beta$ at Earth is then given by
\begin{align}
     \!\!\! f_{\nu_\beta}(E_\nu; \alpha) \! &=\! \sum_i \absVal{U_{fi}}^2 f_{\nu_i}^0(E_\nu; \alpha)
                                     \big[ 1 - P_{\nu_L \to N_R} (\mu_{ii}) \big],
     \label{eq:sn-flux-at-Earth}
\end{align}
where $f_{\nu_i}^0(E_\nu; \alpha)$ is the initial flux emitted from the neutrinosphere, $\mu_{ii}$ is the diagonal magnetic moment for mass eigenstate $\nu_i$, and $P_{\nu_L \to N_R}$ is the conversion probability from $\nu_L$ to $N_R$ in the two-flavour approximation assuming equal masses. Remember that, at the huge matter density inside the supernova core, the mass and flavour bases are aligned: for the normal mass hierarchy we have $f_{\nu_e}^0 = f_{\nu_3}^0$ and $f_{\nu_x}^0 = f_{\nu_2}^0 = f_{\nu_3}^0$, such that $f_{\nu_e} = \cos^2(\theta_{13}) f_{\nu_x}^0 + \sin^2 (\theta_{12}) f_{\nu_e}^0$, where $x = \mu, \tau$.\footnote{We neglect here the fact -- only recently appreciated by the community -- that $\mu$ and $\tau$ neutrinos/anti-neutrinos do not oscillate in exactly the same way \cite{Capozzi:2020kge, Shalgar:2021wlj}. The reasons are the production of muons (but not $\tau$ leptons) in the supernova core \cite{Bollig:2017lki}, as well as the three-flavour vacuum oscillation term in the evolution equation.} Note that, for simplicity, we neglect Earth matter effects, assuming that the supernova occurs in a sky location above the detector.

\begin{figure*}
    \centering
    \begin{tabular}{c@{}c}
        \hspace*{-0.7cm}
        \includegraphics[width=1.1\columnwidth]{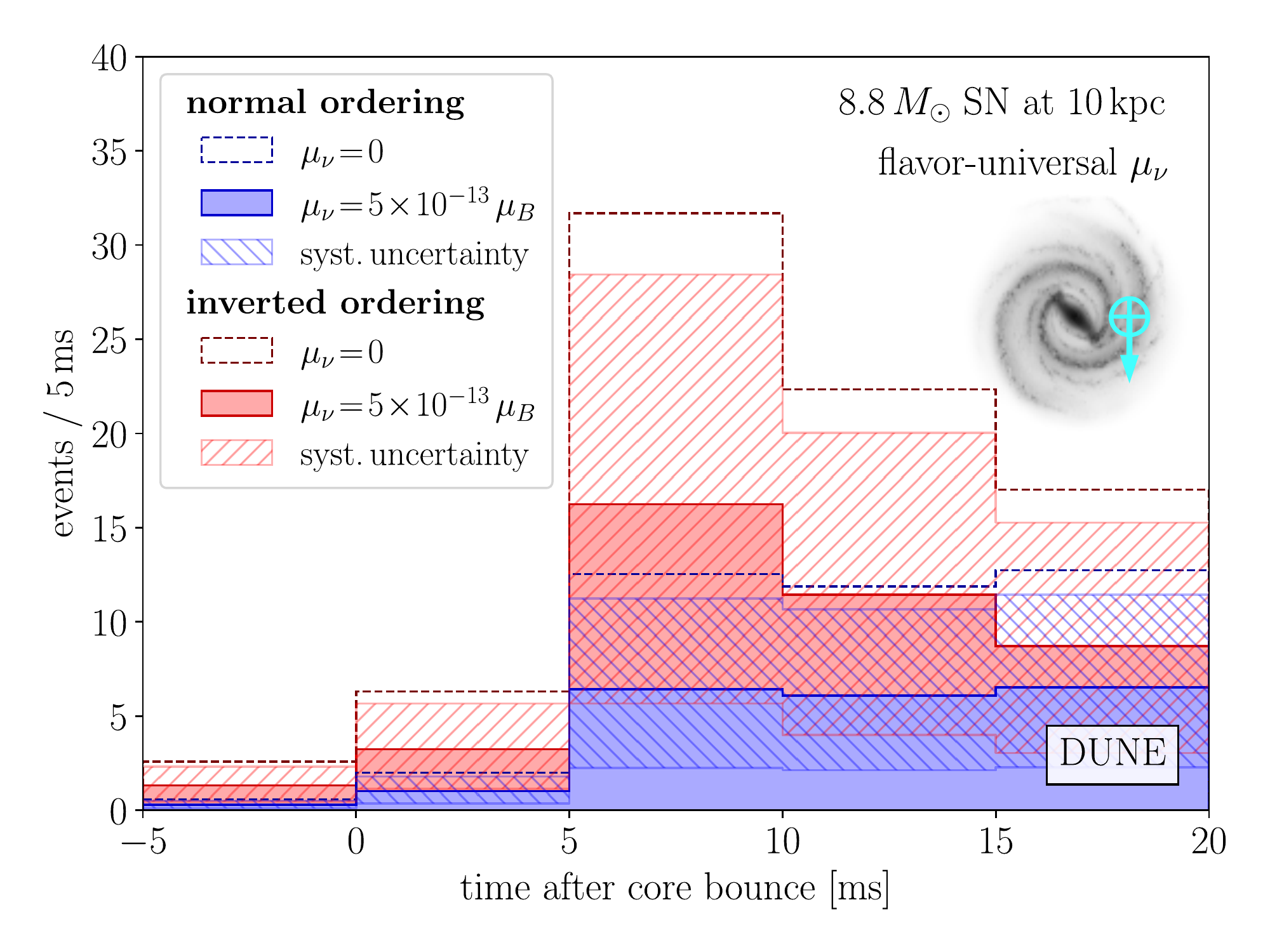}
        &
        \hspace*{-0.2cm}
        \includegraphics[width=1.1\columnwidth]{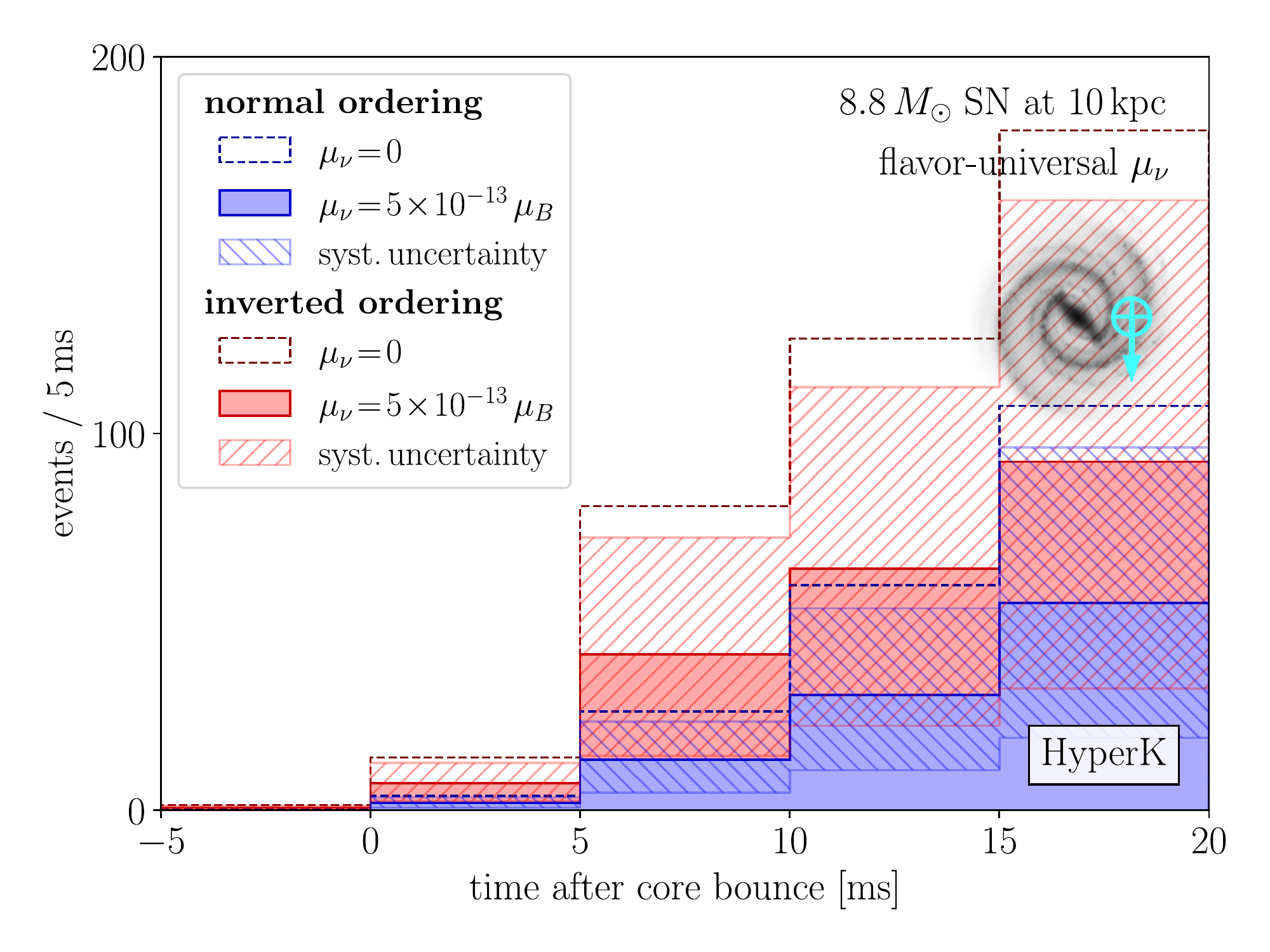}
    \end{tabular}
    \caption{Expected number of events at DUNE (left) and Hyper-Kamiokande (right) from the neutronisation burst of a supernova \SI{10}{kpc} away, both in the standard case without neutrino magnetic moments (dotted histograms), and assuming a Dirac neutrino magnetic moment of $10^{-13} \mu_B$, more than an order of magnitude below the best current limit (solid shaded histograms). We show results assuming both normal (blue) and inverted (red) mass ordering, which determines flavour conversion inside the supernova. Hatched bands indicate the 10\% uncertainty in the initial neutrino flux as well as the uncertainty from our poor knowledge of Galactic magnetic fields. The orientation of the assumed line of sight in the galaxy is indicated by the inset on the right, where the $\oplus$ sign marks our location.}
    \label{fig:sn-event-rates}
\end{figure*}

For a general magnetic field configuration, $P_{\nu_L \to N_R}$ is computed by solving \cref{eq:H2} numerically, using galactic magnetic field profiles generated according to the prescription from \cref{sec:B-field-MW}. \Cref{eq:sn-flux-at-Earth} then yields the expected fluxes of neutronisation burst neutrinos at terrestrial detectors.  The two detectors we consider are Hyper-Kamiokande \cite{Hyper-Kamiokande:2018ofw, Lagoda:2017qaj} and DUNE~\cite{Abi:2020evt, Abi:2020lpk}.  For Hyper-Kamiokande, the total fiducial mass is \SI{374}{kt} (two modules of \SI{187}{kt} fiducial mass each), while for DUNE it will be \SI{40}{kt} (four modules of \SI{10}{kt} fiducial mass each). The most important detection channels in a water \v{C}erenkov detector like Hyper-Kamiokande are inverse beta decay (see refs.~\cite{GilBotella:2003sz,Strumia:2003zx} and more recently ref.~\cite{Ricciardi:2022pru}),
\begin{align}
    \bar{\nu}_e + p \to e^+ + n \,,
    \label{eq:ibd}
\end{align}
and
\begin{align}
    \nu_e (\bar{\nu}_e) + {}^{16}\text{O} \to e^\pm + X \,.
    \label{eq:nu-O16}
\end{align}
In DUNE's liquid argon time projection chambers, the reaction
\begin{align}
    \nu_e + {}^{40}\text{Ar} \to e^- + {}^{40}\text{K}^*
    \label{eq:nu-Ar40}
\end{align}
dominates.  Moreover, neutrino--electron scattering,
\begin{equation}
    \nu_x + e^- \to \nu_x + e^- \,,
    \label{eq:nu-e-scattering}
\end{equation}
can occur in both detectors. It has smaller cross-sections than the nuclear reactions \labelcref{eq:ibd,eq:nu-O16,eq:nu-Ar40}, but offers significantly better angular resolution. (The latter is important for triangulating the location of the supernova in the sky, but is less relevant for setting limits on neutrino magnetic moments.)  The cross-sections for the above reactions can be found in \cite{Strumia:2003zx, Nakazato:2018xkv,Bahcall:1995mm, GilBotella:2003sz}.  As the observable we are interested in is an overall reduction in the neutrino flux, we will combine all detection channels relevant in a given detector into a single event sample. We will also use only one bin in energy.  We will, however, resolve the time-dependence of the signal using five equidistant bins covering the range from $t = -\SI{5}{ms}$ to \SI{20}{ms}, where $t = 0$ is the moment of core bounce. The magnetic moment-induced flux deficit is expected to be the same in all time bins, so this binning serves only to highlight the time evolution of the primary flux and the impact of the neutrino mass ordering on the count rates at Earth.

We repeat the above procedure for 50 different galactic magnetic field profiles, distributed randomly within the uncertainties given in \cref{sec:B-field-MW}. The spread among these 50 different realisations will allow us to gauge the impact of our poor understanding of the Milky Way's magnetic field on the final sensitivity.

A comparison of expected event rates at DUNE and Hyper-Kamiokande with and without magnetic moments is shown in \cref{fig:sn-event-rates} for a galactic magnetic field profile corresponding to the nominal model from \cref{sec:B-field-MW}.  We show results for both the normal (blue) and inverted (red) mass ordering, where the difference between the two cases comes from differences in the adiabatic flavour conversions inside the supernova. Our first observation is that even for $\mu_\nu$ substantially below the current limit ($\SI{1.2e-12}{\mu_B}$, from the position of the tip of the red-giant branch \cite{Capozzi:2020cbu} in the Hertzsprung--Russell diagram), a dramatic deficit of supernova neutrinos can occur. However, due to large magnetic field uncertainties, such a large deficit is not guaranteed. This problem is exacerbated by the fact that, according to \cref{eq:P-LtoR-2f}, the $\nu_L \leftrightarrow N_R$ conversion probability depends sensitively on the exact distance to the supernova. As $P_{\nu_L \to N_R}$ is independent of the neutrino energy in the case of Dirac neutrinos, and since the $\nu_L \leftrightarrow N_R$ oscillation length is much larger than the size of the supernova core, no averaging over oscillations is expected. However, we could be unlucky and it may be located close to a minimum in $P_{\nu_L \to N_R}$. Therefore, as long as only a single supernova is observed and no neutrino deficit is found, it will be difficult to set robust constraints. In other words, the sensitivity of this method is not optimal, while the discovery potential is substantial. If systematic uncertainties in galactic magnetic fields can be reduced to well below our (relatively conservative) error estimates, the sensitivity would dramatically improve and become comparable to the discovery reach.

\subsection{Discovery Reach}
\label{sec:sn-reach}

\begin{figure*}
    \centering
    \begin{tabular}{c@{}c}
        \hspace*{-0.7cm}
        \includegraphics[width=1.1\columnwidth]{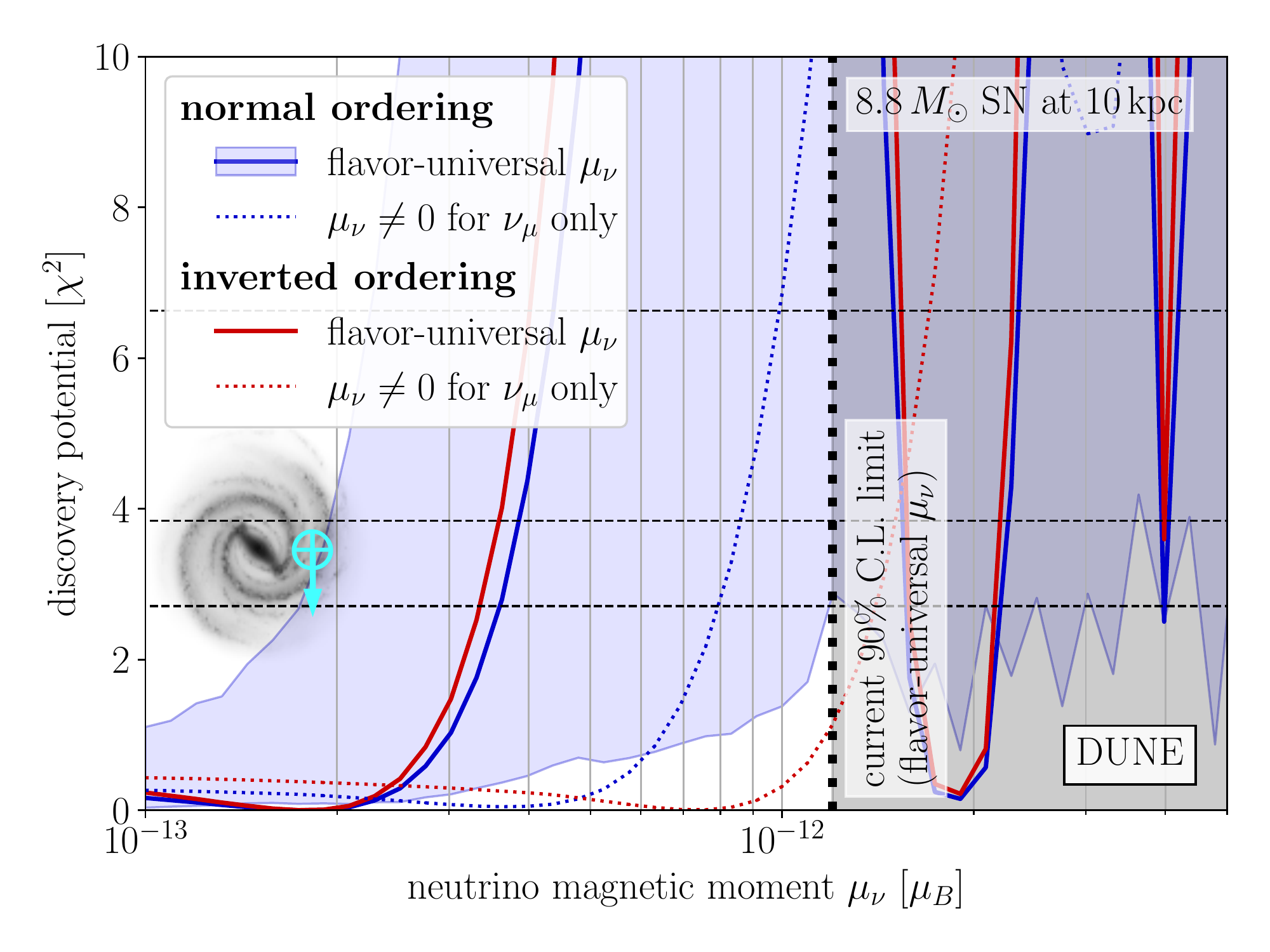}
        &
        \hspace*{-0.2cm}
        \includegraphics[width=1.1\columnwidth]{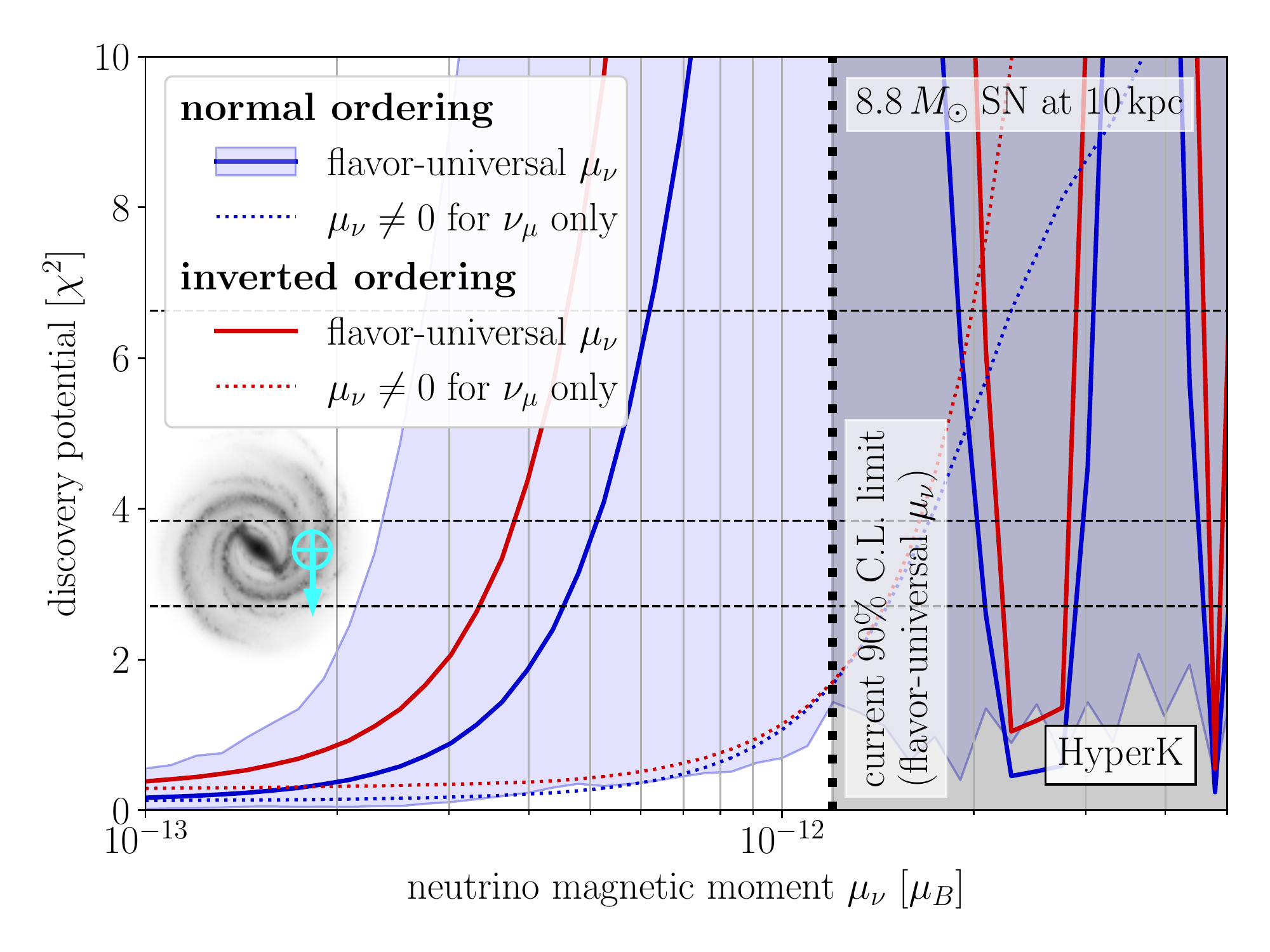}
    \end{tabular}
    \caption{Predicted $\chi^2$ profiles for the analysis of supernova neutronisation burst data in DUNE (left) and Hyper-Kamiokande (right). The assumed true (Dirac) magnetic moment is indicated on the horizontal axis, while the vertical axis shows the level of tension between the simulated ``data'' with $\mu_\nu \neq 0$ and the model prediction for $\mu_\nu = 0$.  We show results for both the normal (blue) and inverted (red) mass ordering, and for flavour-universal magnetic moments (solid) as well as magnetic moments affecting only $\nu_\mu$ (dotted). Lines correspond to the median discovery potential as the assumed true magnetic field model and the true primary flux normalisation are varied within uncertainties. The coloured blue band (shown only for the normal ordering, flavour-universal case for clarity) illustrates the $1\sigma$ variation.  Dashed horizontal lines indicate $\chi^2$ thresholds for one degree of freedom. The region to the right of the vertical dotted line is disfavoured by red-giant cooling constraints \cite{Capozzi:2020cbu}. The orientation of the assumed line of sight in the galaxy is indicated by the inset below the legend, with the $\oplus$ sign marking the location of the Earth.}
    \label{fig:chi-square}
\end{figure*}

The discovery potential of DUNE and Hyper-Kamiokande is best illustrated in \cref{fig:chi-square}, where we show the $\chi^2$ that would be obtained if neutrinos have a non-negligible magnetic moment, but $\mu_\nu = 0$ is assumed in the fit. More precisely, we define
\begin{align}
    \chi^2 &= \underset{a}{\min}
              \sum_i \frac{\big[ n_i(\hat{\mu}_\nu; B, l) - (1+a) \, n_i(0) \big]^2}
                          {(1+a) \, n_i(0)}
            + \frac{a^2}{\sigma_a^2} \,,
    \label{eq:chi2}
\end{align}
where $n_i(\hat{\mu}_\nu; B, l)$ is the assumed observed number of events in the $i$-th time bin for magnetic moment matrix $\hat{\mu}_\nu$, turbulent magnetic field scale $B$, and outer turbulence scale $l$. Similarly, $n_i(0)$ is the fitted number of events for zero magnetic moment (where the magnetic field parameters are irrelevant). The nuisance parameter $a$ parameterises the uncertainty in the primary neutrino flux, and the pull term $a^2/\sigma_a^2$ penalises deviations from the nominal flux. Here we choose $\sigma_a = 0.1$.

We see from \cref{fig:chi-square} that flavour-universal magnetic moments of $\text{few} \times \SI{e-13}{\mu_B}$ should be detectable in both DUNE and Hyper-Kamiokande. In fact the two experiments have rather similar discovery reach. At somewhat larger magnetic moments, the discovery reach goes through several maxima and minima due to the oscillatory behaviour of $P_{\nu_L \to N_R}$ as a function of $\mu_\nu$. \Cref{fig:chi-square} also illustrate the strong dependence of the discovery reach on the true magnetic field profile. Specifically, the coloured bands indicate the $1\sigma$ variation in the $\chi^2$ curves as the magnetic field and the normalisation of the primary neutrino flux are varied. We find very similar discovery reach for the normal (blue) and inverted (red) mass ordering. If only one neutrino flavour -- here $\nu_\mu$ -- carries a sizeable magnetic moment, the discovery reach is worsened by an order of magnitude.

\begin{figure}
    \centering
    \includegraphics[width=\columnwidth]{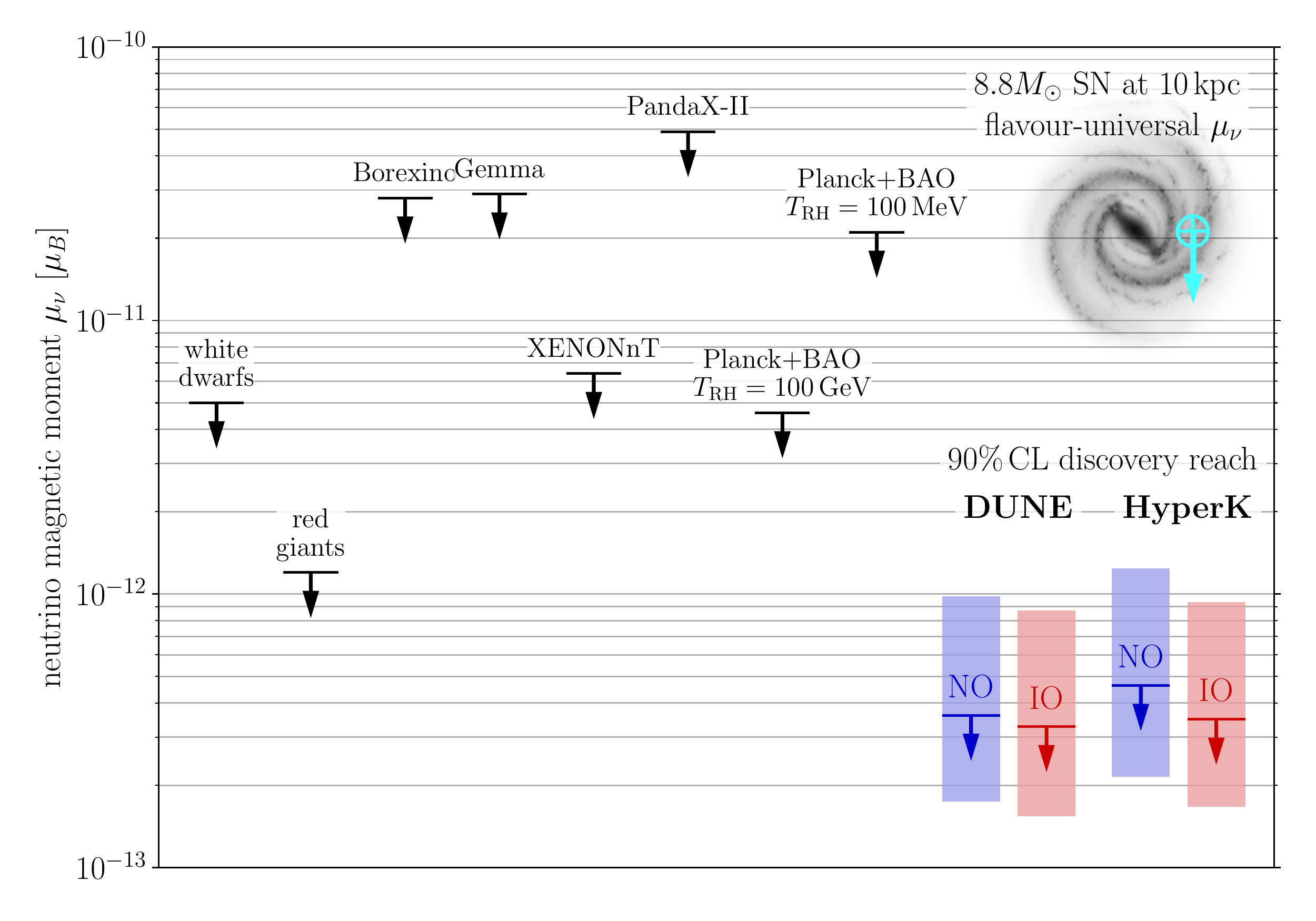}
    \caption{90\% confidence level limits on the Dirac neutrino magnetic moment. The left part of the plot summarises (in black) existing limits from the white dwarf luminosity function \cite{MillerBertolami:2014oki}, the tip of the red giant branch \cite{Capozzi:2020cbu}, solar neutrino observations in Borexino \cite{Borexino:2017fbd}, reactor neutrino measurements in GEMMA \cite{Beda:2013mta}, solar neutrinos in XENONnT \cite{XENONCollaboration:2022kmb} and PandaX \cite{PandaX-II:2020udv}, as well as cosmological $N_\text{eff}$ constraints from the combination of Planck and baryon acoustic oscillation (BAO) data \cite{Carenza:2022ngg, Li:2022dkc}. The coloured lines and bands on the right indicate the anticipated discovery reach in DUNE and Hyper-Kamiokande, determined from the $\chi^2$ curves in \cref{fig:chi-square}. Bands correspond to $1\sigma$ variations of the magnetic field parameters in our model and $1\sigma$ variations of the initial neutrino flux prediction. We have assumed an $8.8 M_\odot$ supernova at a distance of \SI{10}{kpc}, with the line of sight indicated by the small inset on the top right. The $\oplus$ sign in that inset marks the location of the Earth in the Milky Way. The magnetic moments are taken to be flavour-universal.}
    \label{fig:limits}
\end{figure}

In \cref{fig:limits}, we compare the expected discovery reach of DUNE and Hyper-Kamiokande to existing limits on neutrino magnetic moments. For each magnetic field profile, we define the 90\%~CL discovery reach as the value of $\mu_\nu$ for which $\chi^2$ first crosses the threshold of 2.71. We see that even for unfavourable magnetic field configurations, it is likely that the next Galactic supernova should be sensitive to magnetic moments below the best current limit (the one based on the location of the tip of the red giant branch in the Hertzsprung--Russell diagram, which would be modified if $\nu_L \to N_R$ conversions entail extra energy loss from stars). Under favourable circumstances, even an order of magnitude improvement is possible.  The discovery reach is roughly the same for DUNE and Hyper-Kamiokande, and it is moreover largely independent of the neutrino mass ordering. Note that \cref{fig:limits} is for flavour-universal magnetic moments; if only some neutrino flavours experience a large magnetic moment, limits will be correspondingly weaker.

These results show that, thanks to robust predictions of the primary neutrino flux, the neutronisation burst of a Galactic supernova can be a powerful tool to search for physics beyond the Standard Model. They also show that a better understanding of Galactic magnetic fields would greatly aid in this endeavour (besides, of course, being a very interesting goal in itself due to its impact on Galactic astrophysics and radio astronomy).

\section{Ultra-high Energy Neutrinos}
\label{sec:uhe}

The discovery of an ultra-high energy astrophysical neutrino flux at IceCube \cite{IceCube:2013low, IceCube:2014stg} provides us with a new opportunity of probing neutrino properties and observing the high-energy Universe. The neutrino flux can be approximated by a power law
\begin{align}
    \Phi(E) = \phi_0 \times \bigg(\frac{E}{\SI{100}{TeV}} \bigg)^{-\gamma} \,.
\end{align}
For $\SI{25}{TeV} < E < \SI{2.8}{PeV}$, the best-fit parameters are \cite{IceCube:2020wum}
\begin{align}
    \phi_0 = (6.37_{-1.62}^{+1.46}) \times 10^{-18} \,
             \text{GeV}^{-1} \, \text{s}^{-1} \, \text{sr}^{-1} \, \text{cm}^{-2},
    \label{eq:uhe-phi0}
\end{align}
and
\begin{align}
    \gamma = 2.87_{-0.19}^{+0.20} \,.
    \label{eq:uhe-gamma}
\end{align}
Note, however, that there is some tension between different measurements of $\phi_0$ and $\gamma$. For instance, the spectral index quoted here based on IceCube's High Energy Starting Event (HESE) sample, $2.87_{-0.19}^{+0.20}$~\cite{IceCube:2020wum}, is higher than the values extracted from upward-going muons ($2.37 \pm 0.09$~\cite{IceCube:2021uhz}), from cascade-like events ($2.53 \pm 0.07$~\cite{IceCube:2020acn}), and from a study of the inelasticity of high-energy neutrino interactions ($2.62 \pm 0.07$~\cite{IceCube:2018pgc}. As the parameters in \cref{eq:uhe-phi0,eq:uhe-gamma} are extracted from data and attempts to predict them from first principles are fraught with large uncertainties, they cannot be used directly to constrain new physics. In particular, using a flux deficit to probe disappearance of left-handed Dirac neutrinos into invisible right-handed states through the magnetic moment operator, as proposed in \cref{sec:sn} for Galactic supernova neutrinos, is not an option here.  However, if magnetic moments are flavour non-universal, their effect may be observable because the $\nu_L \to N_R$ conversion probability will then be different for different $\nu_L$ flavours. This implies that the flavour composition of the neutrino flux arriving at Earth will be different compared to the Standard Model. Indeed, neutrino telescopes like IceCube and KM3NeT have at least some flavour sensitivity.

Magnetic moment-induced flavour conversions of ultra-high energy astrophysical neutrinos have been studied previously in refs.~\cite{Kurashvili:2017zab, Alok:2022pdn, Lichkunov:2022mjf, SinghChundawat:2022mll} in the context of spin--flavour precession, but not for $\nu_L \to N_R$ conversions.

A simple back-of-the-envelope estimate leads us to expect excellent sensitivity in neutrino telescopes.  Namely, for a source at $\mathcal{O}(\si{Gpc})$ distance and intergalactic magnetic fields of order \si{nG}, the oscillation phase in \cref{eq:P-LtoR-Dirac} approaches unity already for $\mu_\nu \sim 10^{-16} \mu_B$. Taking into account conversion also in the intracluster medium, where magnetic fields are stronger according to the discussion in \cref{sec:B-field-extragalactic}, leads to an even more promising estimate.

\begin{figure*}
    \centering
    \begin{tabular}{cc}
        \includegraphics[width=\columnwidth]{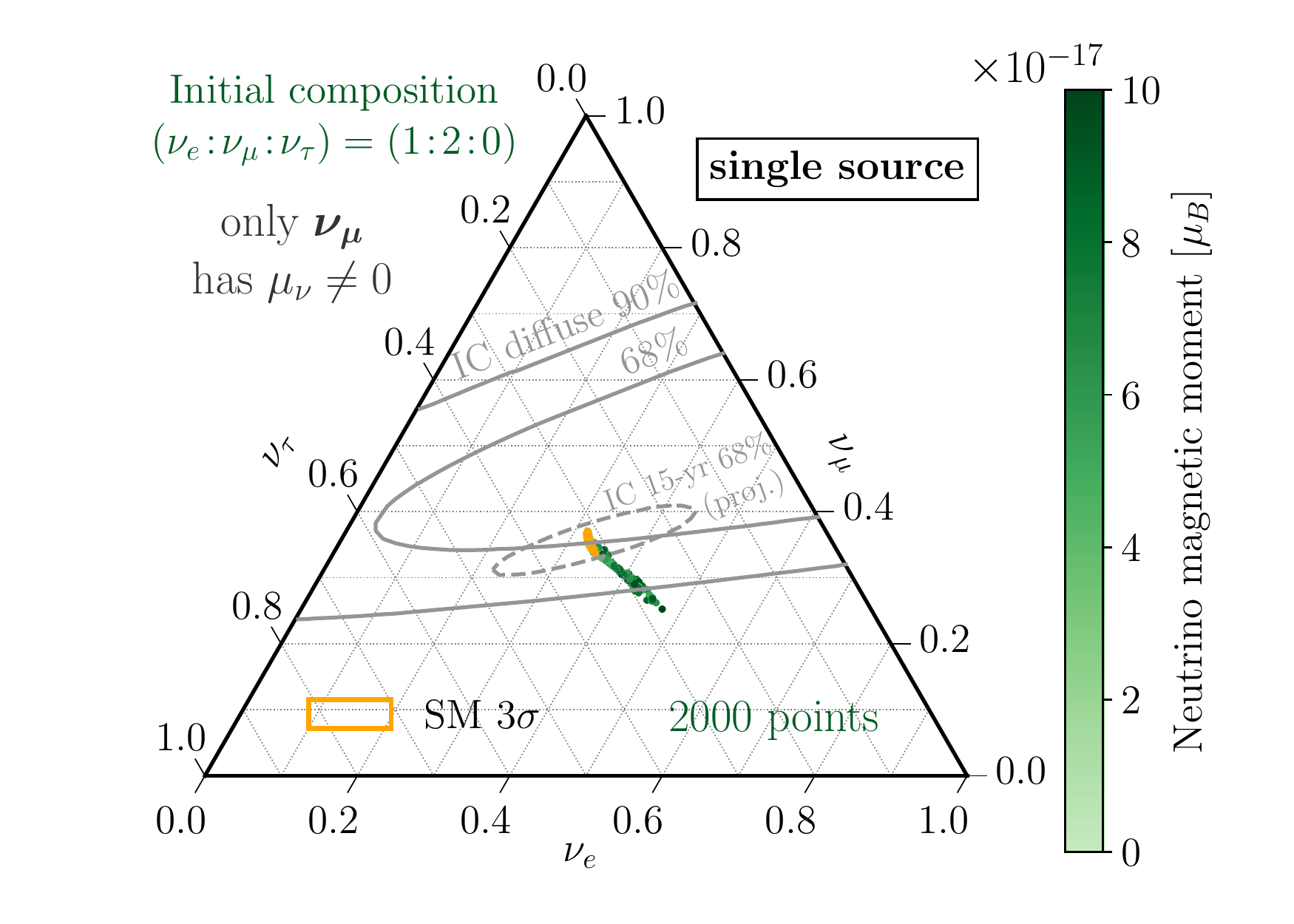} &
        \includegraphics[width=\columnwidth]{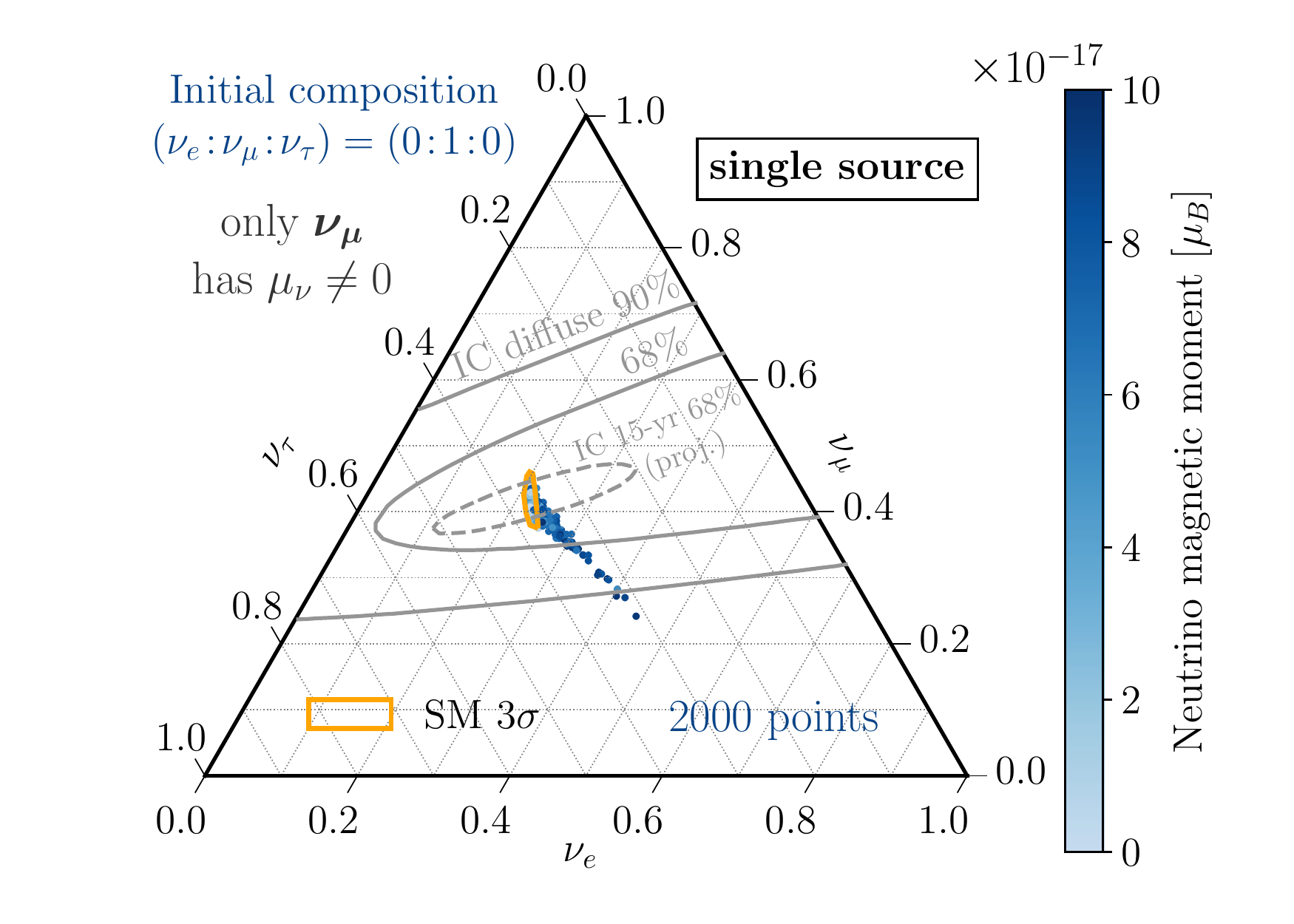}
    \end{tabular}
    \caption{Predicted flavour ratios of astrophysical Dirac neutrinos in the Standard Model (small regions delineated in orange) and in presence of flavour non-universal neutrino magnetic moments (scattered points, with colour code indicating the magnitude of the magnetic moment). For the SM, the uncertainties in the neutrino mixing angles are taken into account, while for non-negligible magnetic moments, also the uncertainty in the extragalactic magnetic fields is included following the procedure outlined in \cref{sec:B-field-extragalactic}. The source is assumed to be between 0.1 and \SI{5}{Gpc} away. We have used a different random magnetic field profile for each point, thus reflecting the range of possible outcomes if only events from a single source are considered.  Expected results for a diffuse flux, and for magnetic moments with a different flavour structure, are shown in \cref{sec:uhe-appendix}. We compare results for a pion decay source (initial flavour ratios $(1:2:0)$, left panel) and for a muon-damped source (initial flavour ratios $(0:1:0)$, right panel). Finally, current constraints (grey solid ellipses) \cite{IceCube:2015gsk} and future sensitivities (grey dashed ellipses) \cite{IceCube-Gen2:2020qha} for IceCube measurements on the diffuse flux are shown as well.}
    \label{fig:flavour-triangles}
\end{figure*}

And indeed, this optimism appears to be justified as illustrated by the ``flavour triangles'' in \cref{fig:flavour-triangles}. Each of the three axes in these plots corresponds to the fractional admixture of one neutrino flavour to the observed flux. The triangular boundaries visualise the requirement that the $\nu_e$, $\nu_\mu$, and $\nu_\tau$ fractions must add up to 1.  We compare the flavour composition at Earth expected for two astrophysically motivated flavour combinations: \emph{(i)} a neutrino flux produced in the decay of high-energy pions in a low-density environment, which would give an initial flavour ratio $(\Phi_{\nu_e} : \Phi_{\nu_\mu} : \Phi_{\nu_\tau}) = (1:2:0)$, see the left panel in \cref{fig:flavour-triangles}; and \emph{(ii)} a flux from a source where pions decay in a dense environment, so that the resulting muons lose a considerable amount of energy before decaying. In this case, secondary muon decays do not contribute to the neutrino flux at high energies, and the initial flavour ratio is $(0:1:0)$ (right panel in \cref{fig:flavour-triangles}). After propagation, these initial fluxes have evolved to the flavour ratios shown as green/blue dots, and the uncertainties in the final flavour ratios due to mixing angle uncertainties are indicated by the orange contours, with angles and uncertainties taken from the NuFit~5.1 fit \cite{Esteban:2020cvm}. As is well known, IceCube's current constraints (large grey elliptical lines from ref.~\cite{IceCube:2015gsk}, see also ref.~\cite{Bustamante:2019sdb}) are not sufficient to resolve these mixing angle uncertainties, while future constraints (grey dashed ellipses, from ref.~\cite{IceCube-Gen2:2020qha}) will get close. Further improvements are expected with the Gen-2 upgrade of the detector.

IceCube could, however, already now detect non-zero neutrino magnetic moments if they violate lepton flavour universality. This is illustrated by the scattered coloured points in \cref{fig:flavour-triangles}. We assume here that only $\nu_\mu$ have a non-vanishing magnetic moment, with each point in the scatter plot corresponding to a random value of the magnetic moment, as indicated by its colour, as well as a random choice of mixing parameters, each of them drawn from a normal distribution with central value and error bar from NuFit~5.1 \cite{Esteban:2020cvm}. Results for a magnetic moment with a different flavour structure, namely a scenario where only the $\nu_2$ mass eigenstate has non-zero $\mu_\nu$, are shown in \cref{sec:uhe-appendix}. In a similar way as we did for supernova neutrinos, we again consider incoherent propagation and independent conversions between left-handed and right-handed neutrinos for each mass state, solving \cref{eq:schroedinger} numerically in the two-flavour approximation and neglecting the vacuum oscillation term in the Hamiltonian.\footnote{The latter assumption is always justified for Dirac or pseudo-Dirac neutrinos. But even for large splitting between $\hat{M}_\nu$ and $\hat{M}_N$ it is justified in the high-energy limit where $B \mu \gg \Delta m_{N1}^2 / (2p)$. In this limit, our results also become independent of energy.} For each of the coloured points in \cref{fig:flavour-triangles}, we have chosen one of 50 extragalactic magnetic field profiles, each of them randomly drawn from the distributions described in \cref{sec:B-field-extragalactic}. Moreover, the distance of the neutrino source for each field profile has been chosen randomly between \SI{100}{Mpc} and \SI{5}{Gpc}.  Each point thus describes a possible outcome of a measurement on a sample of neutrinos from a \emph{single} source. (We comment below and in \cref{sec:uhe-appendix} on the implications of these assumptions.)  We do not account for possible additional $\nu_L \leftrightarrow N_R$ conversion inside the source due to our complete ignorance of the source magnetic field. Doing so is conservative because additional conversion would likely increase the deviation from the Standard Model point in the flavour triangle and thus increase the experimental sensitivity to neutrino magnetic moments.

\Cref{fig:flavour-triangles} shows that, if the magnetic field configuration along the line of sight is favourable, even tiny lepton flavour universality-violating magnetic moments of order $\text{few} \times 10^{-17} \mu_B$, well below the current limits of $\text{few} \times 10^{-12} \mu_B$, can lead to flavour ratios that deviate significantly from those expected in the Standard Model. Typically, we will have very little information on the magnetic fields neutrinos encounter during propagation, therefore, the limit that can be set in absence of a discovery will be much weaker than $10^{-17} \mu_B$. A possible game-changer could be detailed measurements of Faraday rotation in radio emission from an identified neutrino source. Such a study could significantly constrain the magnetic field profile along the line of sight, thereby reducing the main systematic uncertainty limiting the sensitivity to neutrino magnetic moments.

Our conclusions would remain unchanged if the magnetic moment matrix had several non-zero entries, as long as these entries all differ by $\mathcal{O}(1)$ factors. Only for close to flavour-universal magnetic moments, deviations from Standard Model expectations would become weaker. Note that, similar to what we observed in \cref{fig:chi-square}, the conversion probability for neutrino magnetic moments well above the sensitivity limit goes through a series of maxima and minima. We refer the reader to \cref{sec:uhe-appendix} for a brief discussion of alternative magnetic moment matrices.

Let us emphasise again that the large deviations from Standard Model expectations we predict here are unique to analyses using only neutrinos from a single source (that is, a single line of sight and magnetic field profile).  To predict the flavour ratios of a diffuse flux, we would need to average over many lines of sight, which would lead to wash-out: each mass eigenstate experiences on average about 50\% disappearance, leaving the flavour ratios unchanged. Results for a diffuse neutrino flux are shown in \cref{fig:flavour-triangles-multi-source_mu_only,fig:flavour-triangles-nu1_only} in \cref{sec:uhe-appendix}.

We should of course keep in mind that obtaining flavour information on neutrinos from a point source is very challenging. Only muon neutrinos can be observed with sufficient angular resolution to associate them with a particular source based on the sky location they are coming from. For $\nu_e$ and $\nu_\tau$, such an association is only possible based on timing. In other words, the source must be transient. To date, IceCube has identified two point sources of high-energy neutrinos, the blazar TXS~0506+056 with $\sim 15$ detected neutrinos \cite{IceCube:2018dnn, IceCube:2018cha} and the active nucleus of the NGC~1068 galaxy with about 80 neutrinos \cite{IceCube:2022der}. Both analyses are based on muon tracks only, so the detailed flavour structure of the neutrino flux from these sources is as yet unknown. Moreover, NGC~1068 is relatively close (\SI{14.4}{Mpc}), so its neutrinos will experience much less $\nu_L \leftrightarrow N_R$ conversion than those from a source at $\mathcal{O}(\si{Gpc})$, as we have assumed in \cref{fig:flavour-triangles}. Therefore, NGC~1068 is not optimal for constraining neutrino magnetic moments. TXS~0506+056, on the other hand, is \SI{1.75}{Gpc} away from Earth, and is transient. Radio emission has been observed from both sources, but to the best of our knowledge has never been used to constrain the intermittent extragalactic magnetic fields.

\section{Discussion}
\label{sec:discussion}

In the following, we discuss several generalisations and extensions of the results obtained in the previous sections.

\subsection{Neutrino Magnetic Moments in the Standard Model}
\label{sec:sm}

We have found in \cref{sec:uhe} that flavour ratios of high-energy astrophysical neutrinos may offer superior discovery reach to flavour non-universal neutrino magnetic moments down to $\text{few} \times \SI{e-17}{\mu_B}$. This raises the question if, under favourable circumstances, even the neutrino magnetic moments predicted in the Standard Model (see \cref{eq:munu-sm}) could be within reach. Note that, because of their neutrino mass dependence, these magnetic moments are flavour non-universal, fulfilling a first necessary condition for being detectable using flavour ratios of ultra-high energy neutrinos. Second, according to the sensitivity projections shown in \cref{fig:flavour-triangles} above, future IceCube (and even more so IceCube-Gen2) analyses should be able to detect $\mathcal{O}(10\%)$ changes in the flavour ratios of high-energy cosmic neutrinos, especially once the uncertainties in the standard oscillation parameters will have been further reduced by the next generation of long-baseline experiments. This means that oscillation probabilities $P_{\nu_L \to N_R} \sim \mathcal{O}(10\%)$ are required to make a detection. For magnetic moments of order $\text{few} \times \SI{e-20}{\mu_B}$, this would require
\begin{align}
    \int_0^L \! dx \, B_\perp(x) \simeq \SI{10}{Gpc \, \mu G} \,
    \label{eq:sm-detectability-condition}
\end{align}
according to \cref{eq:P-LtoR-varying-B-1}.  This estimate is valid in case the orientation of the magnetic field does not change along the line of sight. We expect any change in the magnetic field direction to reduce the $\nu_L \leftrightarrow N_R$ oscillation probability. As we have discussed in \cref{sec:B-field-extragalactic}, most observations indicate that extragalactic magnetic fields are $\lesssim \SI{10}{nG}$, while a few point towards larger fields of order 0.2--$\SI{0.6}{\mu G}$ \cite{Kim:1989, Brown:2011, Kronberg:2007wa}. This would be enough to satisfy \cref{eq:sm-detectability-condition} for a source more than \SI{10}{Gpc} away from Earth. (For comparison, recall TXS~0506+056 is at a distance of \SI{1.75}{Gpc}.) Alternatively, one could imagine a line of sight that passes through many galaxy clusters (with intracluster fields of order $\si{\mu G}$), such that the integrated distance neutrinos travel inside clusters is several \si{Gpc}. In any case, the magnetic fields the neutrinos encounter on their way to Earth would need to be oriented in the same direction to avoid cancellations in the integral on the left-hand side of \cref{eq:sm-detectability-condition}. For the same reason, neutrinos should also come from a single point source, corresponding to a single line of sight.

To summarise, neutrino magnetic moments as small as in the Standard Model might be detectable in IceCube if
\begin{enumerate}
    \item a sizeable sample of neutrinos from a single point source at $\sim \SI{10}{Gpc}$ distance is available to enable detection of $\mathcal{O}(10\%)$ deviations in the flavour ratios;

    \item magnetic fields between galaxy clusters are as large as suggested in refs.~\cite{Kim:1989, Brown:2011, Kronberg:2007wa} (in conflict with other constraints), or if an $\mathcal{O}(1)$ fraction of the line of sight passes through galaxy clusters;

    \item magnetic fields along the line of sight are oriented in similar directions to avoid cancellation.
\end{enumerate}

\subsection{Heavier Right-Handed Neutrinos and Decoherence}
\label{sec:coherence}

In \cref{sec:sn,sec:uhe}, we have assumed magnetic moment transitions between the left-handed and right-handed partners of a Dirac pair, that is, we have taken $\Delta m_{N\nu}^2 = 0$. As discussed in \cref{sec:mm}, this is justified as long as $\mu_\nu B_\perp \gg \Delta m_{N\nu}^2/(2p)$.  For larger right-handed neutrino masses, the oscillation amplitude becomes very small according to \cref{eq:P-LtoR-2f}, so this case is phenomenologically less relevant. In particular, the expected discovery reach for magnetic moment-induced transitions to heavier right-handed neutrinos using the methods discussed in this paper is significantly worse than the one for $\Delta m_{N1}^2 \simeq 0$.  Constraints on heavier $N_R$ can still be obtained, but these are typically based on inelastic $\nu_L \to N_R$ transitions \cite{Vogel:1989iv}, followed by $N_R$ decay \cite{Brdar:2020quo}.

This raises an interesting theoretical question. Namely, when can $\nu_L \leftrightarrow N_R$ transitions be described as an interference phenomenon similar to conventional neutrino oscillation (as we have done in this paper so far), and when is a description as a hard scattering process more appropriate? The key to answering this question is the Heisenberg principle. If the $\nu_L$ and $N_R$ masses are sufficiently similar so that the small changes in energy and momentum incurred during a $\nu_L \leftrightarrow N_R$ transition fall below the quantum mechanical energy and momentum uncertainties, neutrino interactions with the magnetic field are coherent and oscillations can occur. This is no longer the case if the $N_R$ are significantly heavier than the active neutrinos. Then it is at least in principle possible based on kinematics to determine where along the neutrino trajectory the conversion has happened. In this regime, the transition probability is most easily obtained by evaluating the Feynman diagram describing the inelastic scattering of a neutrino on one of the particles that source the magnetic field. We will denote the two scenarios the ``coherent'' and ``incoherent'' scattering regimes, respectively.

The natural question to ask in this context is how and where the transition between these two very different theoretical descriptions occurs. We will here focus on the ``where'' -- that is, the threshold value of $m_N$ that separates the two regimes. The ``how'' -- that is, the detailed dynamics of the transition, is discussed in \cref{sec:coherence-calculation}.

Here we proceed with a simple derivation of the $m_N$ threshold. Consider inelastic scattering $\nu_L \to N_R$ on a hypothetical massive target at rest, such that no energy transfer occurs. We compare the momentum transferred to the target, $\Delta p$, in this hypothetical process to the Heisenberg momentum uncertainty, $\sigma_p$, of the electrons in the interstellar medium (ISM) which source interstellar magnetic fields. If $\Delta p > \sigma_p$, the process $\nu_L + e \to N_R + e$ leads to an appreciable change in the electron's quantum state, so the scattering will be incoherent. In contrast, if $\Delta p < \sigma_p$, the electron wave function remains largely unaffected, and the $\nu_L \to N_R$ flavour conversion is well-described as a coherent oscillation process using the formalism from \cref{sec:mm}.

If we denote the initial and final neutrino momenta by $\vec{k}_\nu$ and $\vec{p}_N$, respectively, we have
\begin{align}
    \!\!\Delta p &= |\vec{p}_N - \vec{k}_\nu|
              \!=\! \sqrt{2 E_\nu^2 - m_N^2 + 2 E_\nu \sqrt{E_\nu^2 + m_N^2}}.
    \label{eq:Delta-p}
\end{align}
Here, we have used that the $\nu_L$ and $N_R$ energies are identical in our hypothetical scattering process on a very massive target, and we have considered the scattering angle to be $180^\circ$, corresponding to the kinematic configuration with the maximal momentum transfer.  From \cref{eq:Delta-p}, we can immediately derive that the coherence condition, $\Delta p < \sigma_p$, corresponds to the requirement
\begin{align}
    m_N &< 2 \sqrt{\sigma_p (E_\nu - \sigma_p)}
         \simeq 2 \sqrt{\sigma_p E_\nu} \notag\\
        &= \SI{6.3}{keV} \times \bigg( \frac{\SI{1}{cm}}{\sigma_x} \bigg)
                               \bigg( \frac{E_\nu}{\SI{1}{TeV}} \bigg) \,.
    \label{eq:coherence-cond}
\end{align}
In the last step, we have defined the spatial localisation of the ISM electrons $\sigma_x = 1/(2\sigma_p)$.  The benchmark value $\sigma_x \simeq \SI{1}{cm}$ has been chosen based on typical ISM densities. (Keep in mind, though, that the latter can vary by several orders of magnitude.)

\Cref{eq:coherence-cond} shows that the coherent formalism employed in this paper is mostly relevant for fairly light $N_R$.  Only then, $\nu_L \to N_R$ scattering occurs coherently on all electrons along the neutrino trajectory.  The constraint is fairly tight for supernova neutrinos ($E_\nu \sim \SI{10}{MeV}$). But even for the highest-energy cosmic neutrinos with $E_\nu \sim \si{PeV}$, coherent flavour conversion is relevant only for $N_R$ masses well below \SI{1}{MeV}.

\section{Conclusions}
\label{sec:conclusions}

In summary, we have examined the role neutrino magnetic moments play in the propagation of astrophysical neutrinos through both Galactic and extragalactic magnetic fields. 
The key take-home messages are as follows:
\begin{itemize}[leftmargin=*]
    \item (Extra-)Galactic magnetic fields are a powerful discovery tool for neutrino magnetic moments inducing transitions from left-handed active to right-handed neutrinos, $\nu_L \to N_R$. However, this conversion probability is only sizeable for small mass-splitting, namely for the case where the active neutrinos are Dirac or pseudo-Dirac in nature, see \cref{eq:P-LtoR-2f}.
    \item For supernova neutrinos the precise prediction of the flux from the neutronisation burst affords the ability to construct sensitive observables based on the depletion of the observed neutrino flux through $\nu_L \to N_R$ conversions in our Galactic magnetic field, see \cref{fig:sn-event-rates}. However, due to large uncertainties on both the large-scale coherent and small-scale turbulent components of this field (see \cref{sec:B-field-MW}) it is challenging to place robust constraints. Nevertheless, under conservative assumptions on these uncertainties this method affords a discovery reach an order of magnitude superior to current constraints (see \cref{fig:limits}) depending also on the line-of-sight to the supernova. Interestingly these results are largely independent of the details of the supernova magnetic field as the initial neutrino flux before propagation depends only on the ordering of the neutrino mass hierarchy.
    \item An alternative approach that yields potentially far greater sensitivities utilises high-energy astrophysical neutrinos observed in neutrino telescopes. As the sources of these neutrinos are not well understood the sensitivity here arises from deviations in the flavour composition. This method is exceptionally powerful if there is a large hierarchy in the neutrino magnetic moments of various neutrino flavours, and if a sample of neutrinos from a single point source is considered. In this case, probing magnetic moments as small as $10^{-17}~\mu_B$ seems feasible, see \cref{fig:flavour-triangles}. Measuring the flavour composition of neutrinos from a point source would most likely require this source to be transient due to the limited angular resolution of $\nu_e$ and $\nu_\tau$ reconstruction.  But even for diffuse astrophysical neutrino fluxes, the discovery reach is still several orders of magnitude beyond current limits.
    \item Under very favourable circumstances neutrino magnetic moments arising purely from Standard Model contributions may be detectable in IceCube, see \cref{sec:sm}. This however will require an extremely distant point source, with not only favourable orientation of the magnetic field (to avoid cancellations) but also an $\mathcal{O}(1)$ fraction of the line-of-sight passing through galaxy clusters or other structures with enhanced magnetic fields. 
    \item Lastly, we have discussed for which values of the $\nu_L$--$N_R$ mass-splittings our formalism which describes $\nu_L \leftrightarrow N_R$ transitions as coherent oscillations is valid. In particular, it is required that the magnetic field act coherently on the propagating neutrinos. In \cref{sec:coherence} we sketched an argument showing that for ultra-high-energy neutrinos this requires sub-MeV $N_R$ masses, while in \cref{sec:coherence-calculation} a detailed calculation of this decoherence criterion is presented. 
\end{itemize}
To conclude, we have established several new and highly promising avenues towards discovering tiny neutrino magnetic moments in theories beyond the Standard Model -- or even in the Standard Model itself.  So far, these methods are limited by our poor understanding of Galactic and intergalactic magnetic fields, but we note that there is significant room for improvement in this field, for instance through observations of Faraday rotation in fast radio bursts and through large-scale radio surveys. Even modest advances in our knowledge about the magnitude and structure of magnetic fields will allow for tremendous improvements in the sensitivity and discovery reach for neutrino magnetic moments.

\section*{Acknowledgements}

It is a pleasure to thank Vedran Brdar and Admir Greljo for inspiring and useful discussions in the early stages of this work. We also thank Evgeny Akhmedov for correcting a statement regarding resonant spin-flavour oscillations in the Dirac case. EW's work is supported by the Collaborative Research Center SFB~1258 of the German Research Foundation.  The plots in this paper have been created using \texttt{Matplotlib} (v3.5.2) \cite{Hunter:2007} as well as the Python \texttt{ternary} package (v1.0.8) \cite{Harper:2015}.

\appendix
\section{Line-of-Sight Dependence of Galactic $\nu_L \to N_R$ Conversion}
\label{sec:los}

\begin{figure*}
    \centering
    \begin{tabular}{c@{}c}
        \hspace*{-0.7cm}
        \includegraphics[width=1.1\columnwidth]{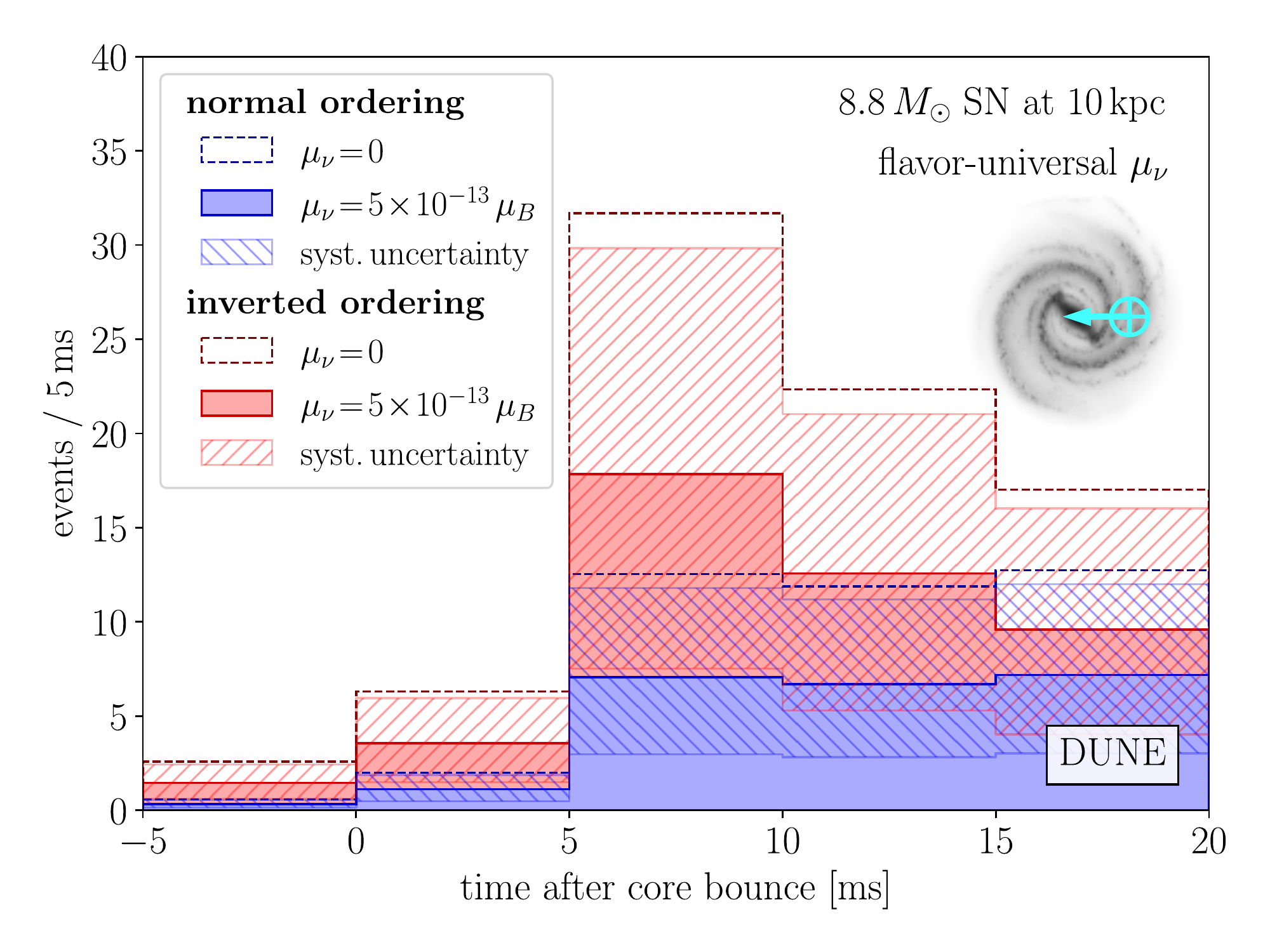}
        &
        \hspace*{-0.2cm}
        \includegraphics[width=1.1\columnwidth]{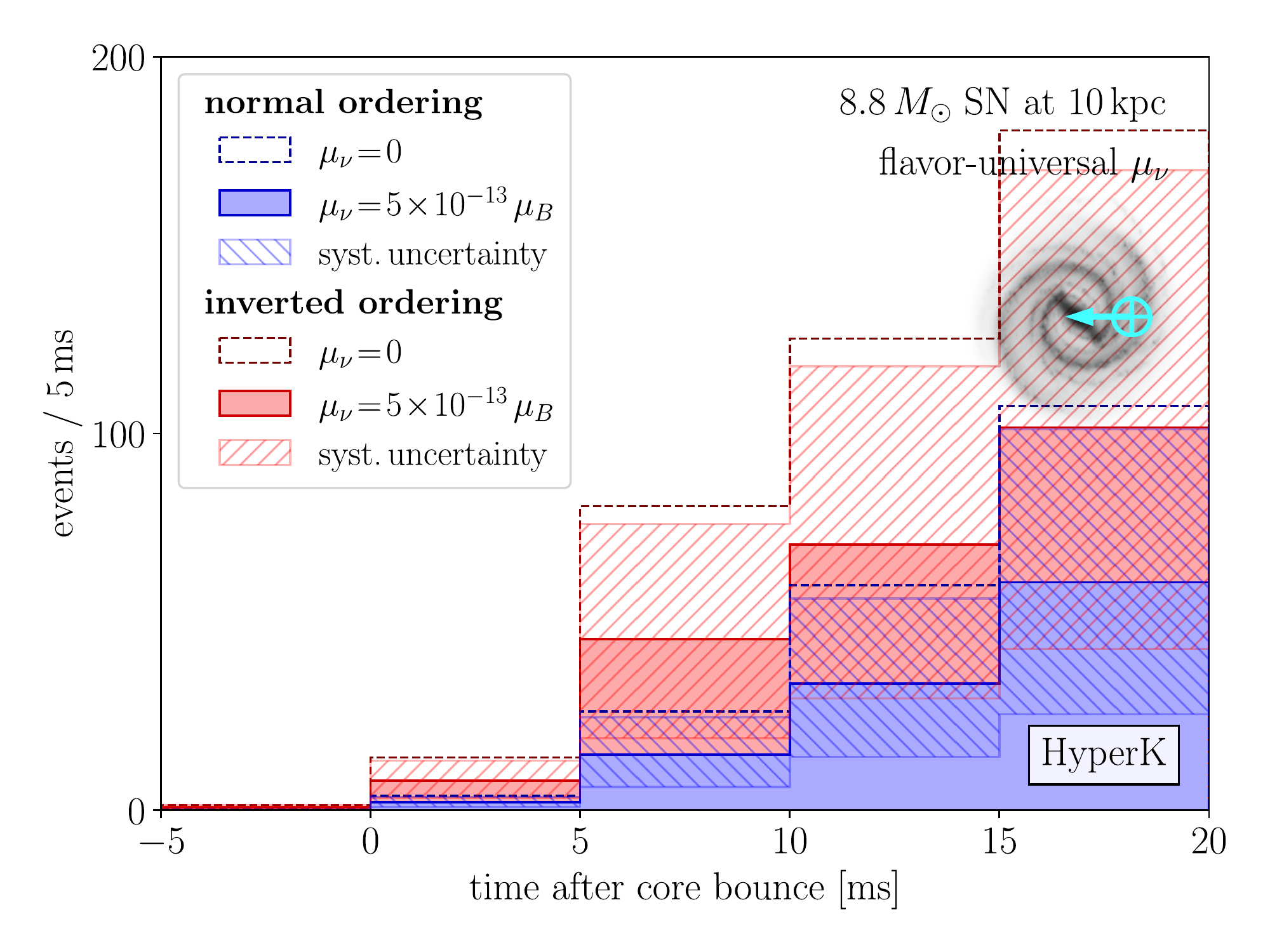}
        \\
        \hspace*{-0.7cm}
        \includegraphics[width=1.1\columnwidth]{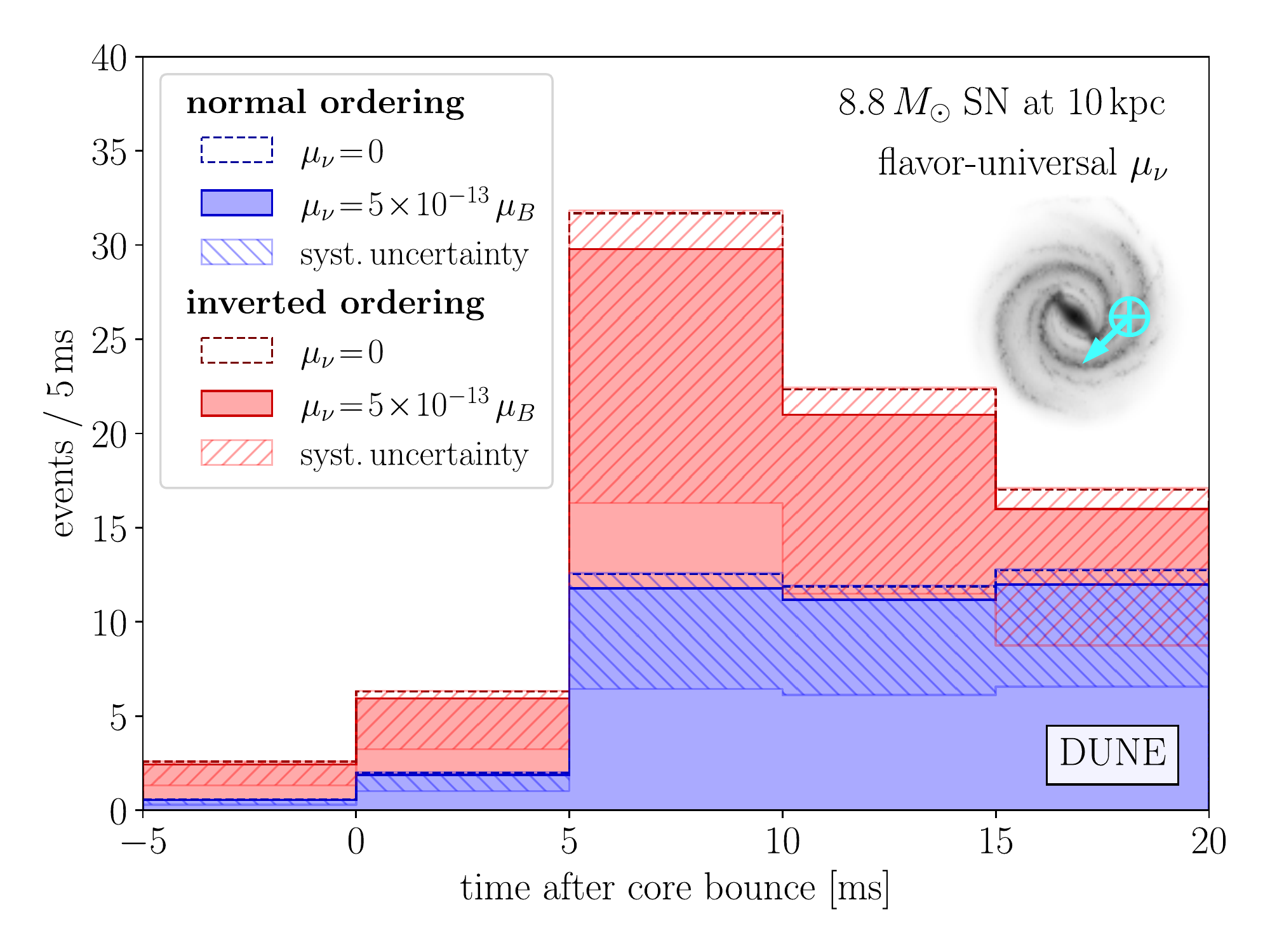}
        &
        \hspace*{-0.2cm}
        \includegraphics[width=1.1\columnwidth]{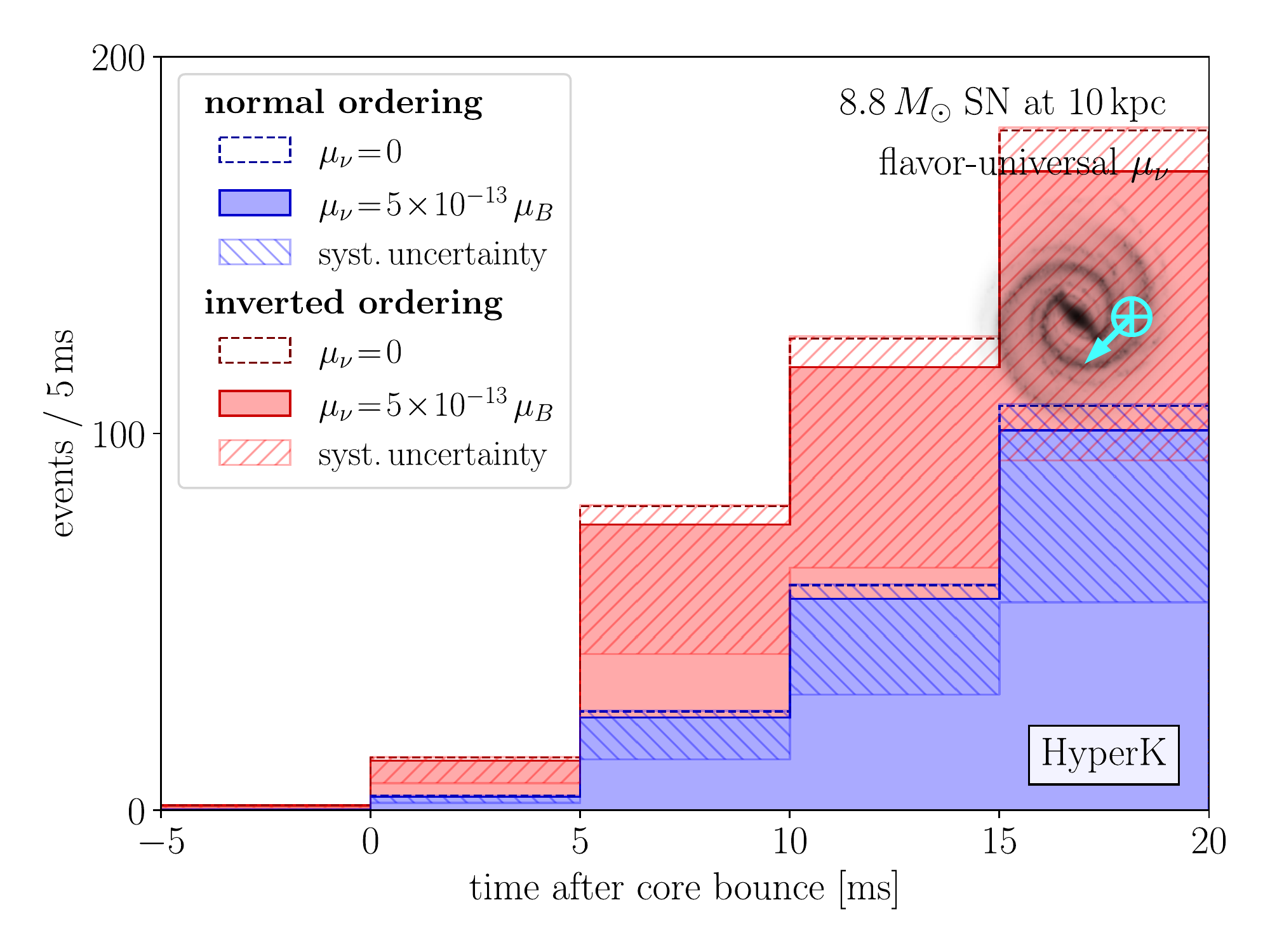}
    \end{tabular}
    \caption{Similar to \cref{fig:sn-event-rates}, we show the expected number of events at DUNE (left) and Hyper-Kamiokande (right) from the neutronisation burst of a supernova with and without magnetic moment-induced $\nu_L \leftrightarrow N_R$ conversion. Compared to \cref{fig:sn-event-rates}, we here vary the assumed location of the supernova in the Milky Way, as indicated by the inset on the right.}
    \label{fig:sn-event-rates-appendix}
\end{figure*}

In \cref{sec:sn} we have illustrated that a substantial fraction of active neutrinos emitted by a supernova can be converted to invisible $N_R$ if non-negligible magnetic moments allow for transitions between the two states in the Galactic magnetic fields. We have illustrated this point in \cref{fig:sn-event-rates} for one particular line of sight. As the coherent component of the Galactic magnetic field strongly depends on the line of sight, we show in \cref{fig:sn-event-rates-appendix} sample spectra for several other lines of sight. We confirm that, indeed, there is a strong dependence on the direction from which neutrinos arrive.

\section{Flavour Ratios of Ultra-High Energy Neutrinos: Alternative Scenarios}
\label{sec:uhe-appendix}

\begin{figure*}[p]
    \centering
    \begin{tabular}{cc}
        \includegraphics[width=\columnwidth]{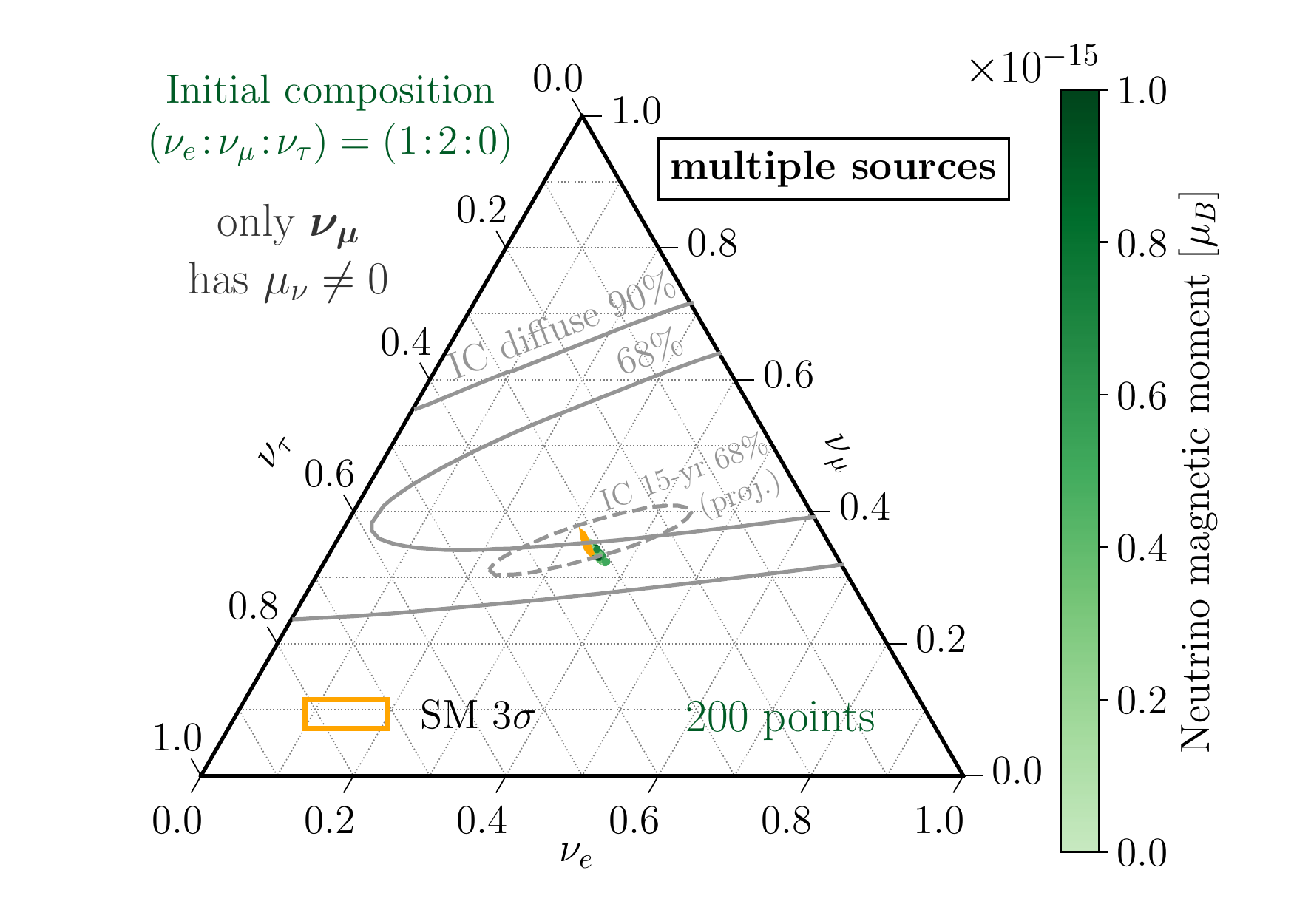} &
        \includegraphics[width=\columnwidth]{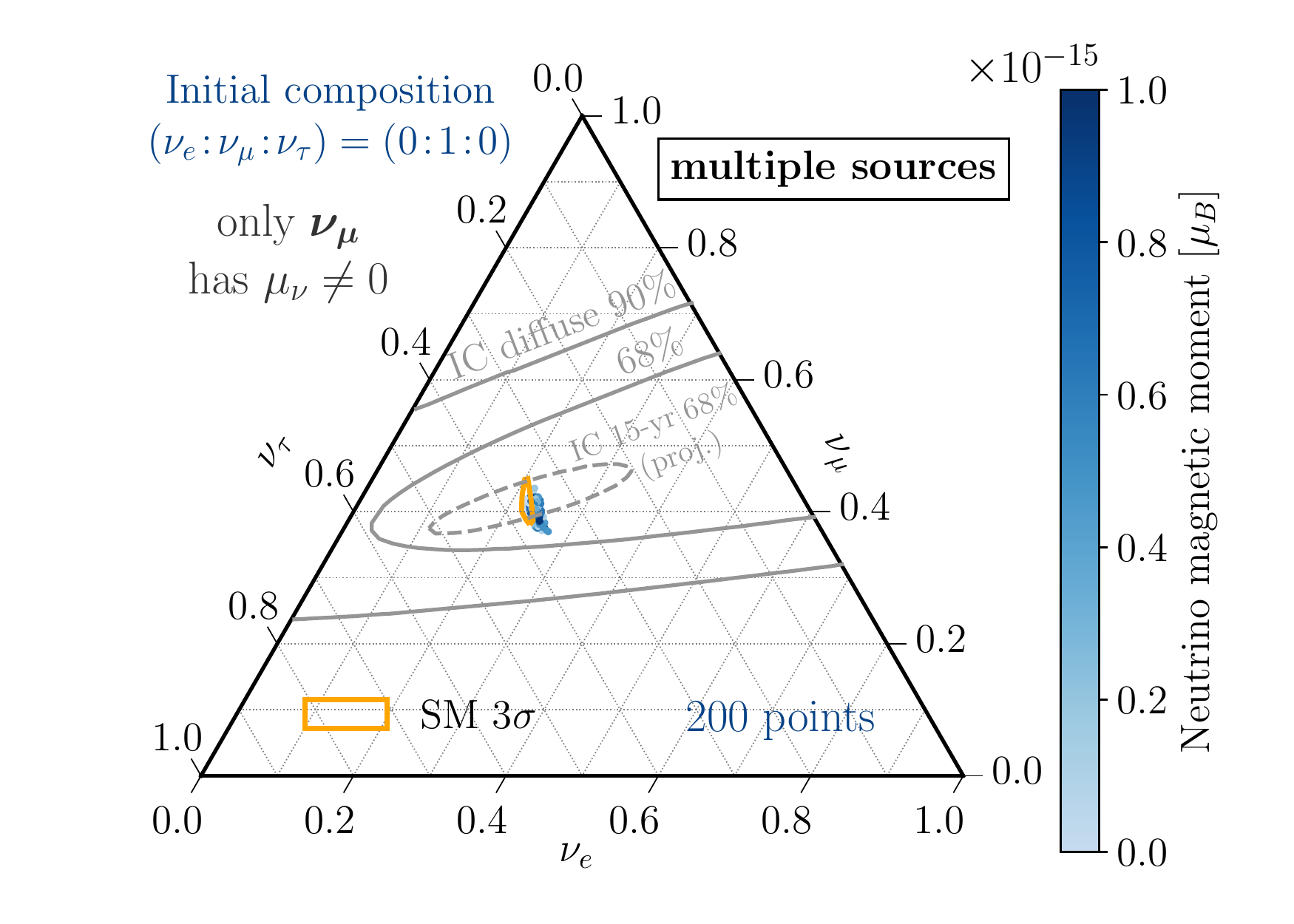}
    \end{tabular}
    \caption{In analogy to \cref{fig:flavour-triangles}, we show here the predicted flavour ratios of astrophysical neutrinos in the presence of a non-zero neutrino magnetic moment affecting only the $\nu_\mu$ flavour. In contrast to \cref{fig:flavour-triangles}, we here assume a diffuse neutrino flux (as opposed to a single point source) and hence average the oscillation probabilities over 50 randomly generated magnetic field profiles. We compare results for a pion decay source (initial flavour ratios $(1:2:0)$, left panel) and for a muon-damped source (initial flavour ratios $(0:1:0)$, right panel). The orange regions correspond to Standard Model predictions, taking into account uncertainties in the neutrino mixing angles. Current and future IceCube constraints for the diffuse neutrino flux \cite{IceCube:2015gsk, IceCube-Gen2:2020qha} are shown in grey. Note the different scaling of the colour bar compared to \cref{fig:flavour-triangles}.}
    \label{fig:flavour-triangles-multi-source_mu_only}
\end{figure*}

To complement the discussion in \cref{sec:uhe}, we show in \cref{fig:flavour-triangles-multi-source_mu_only} the predicted flavour ratios of ultra-high energy astrophysical neutrinos for a diffuse neutrino flux. As in \cref{fig:flavour-triangles}, we focus first on the case where only $\nu_\mu$ carry a non-zero magnetic moment. As anticipated in \cref{sec:uhe}, deviations from the Standard Model are smaller in this case because a measurement on the diffuse flux averages over many different lines of sight with randomly varying magnetic field profiles and therefore randomly varying $\nu_L \to N_R$ conversion probabilities. Note the different scaling of the colour bar here compared to \cref{fig:flavour-triangles}.

\begin{figure*}[p]
    \centering
    \begin{tabular}{cc}
        \includegraphics[width=0.9\columnwidth]{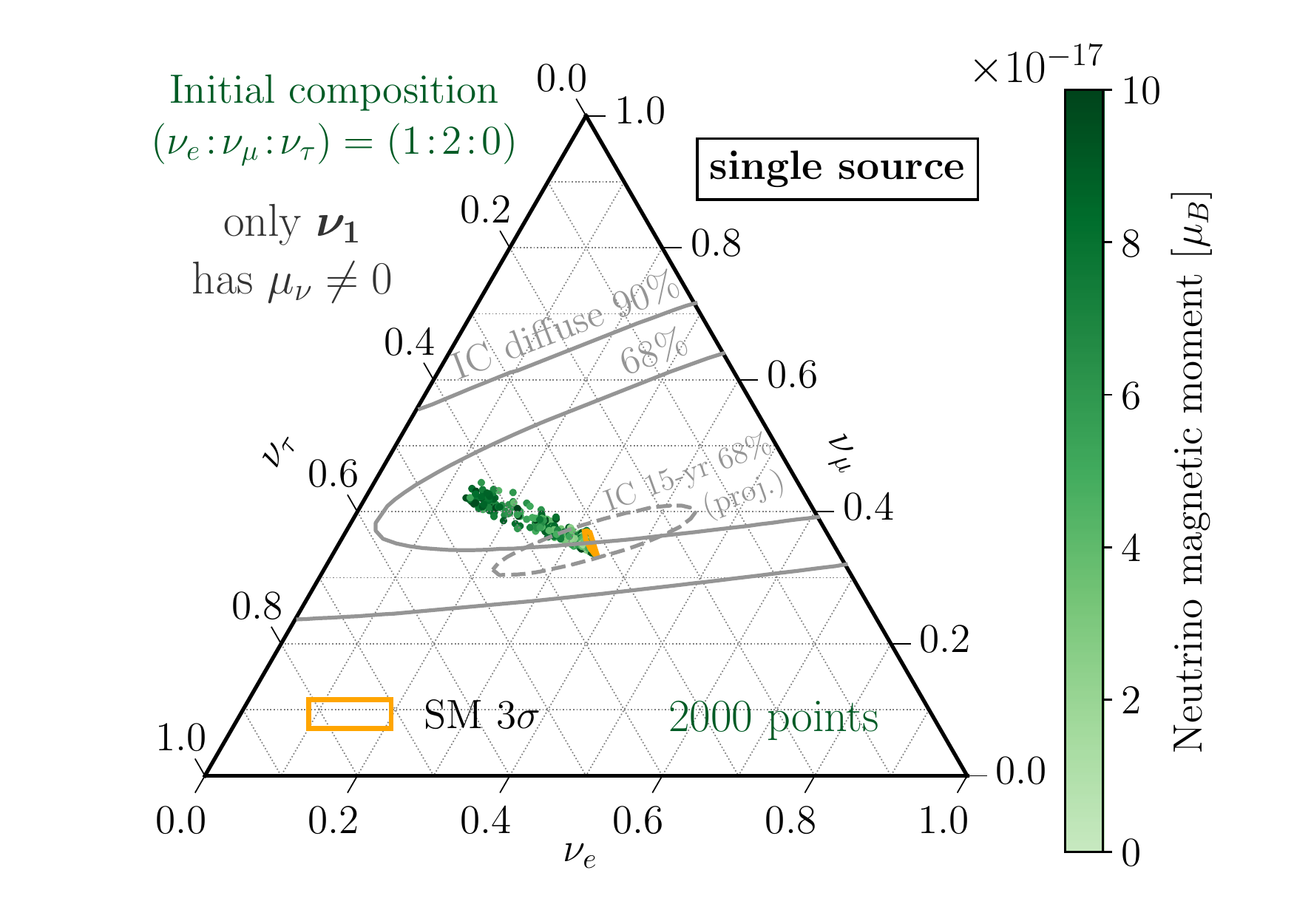} &
        \includegraphics[width=0.9\columnwidth]{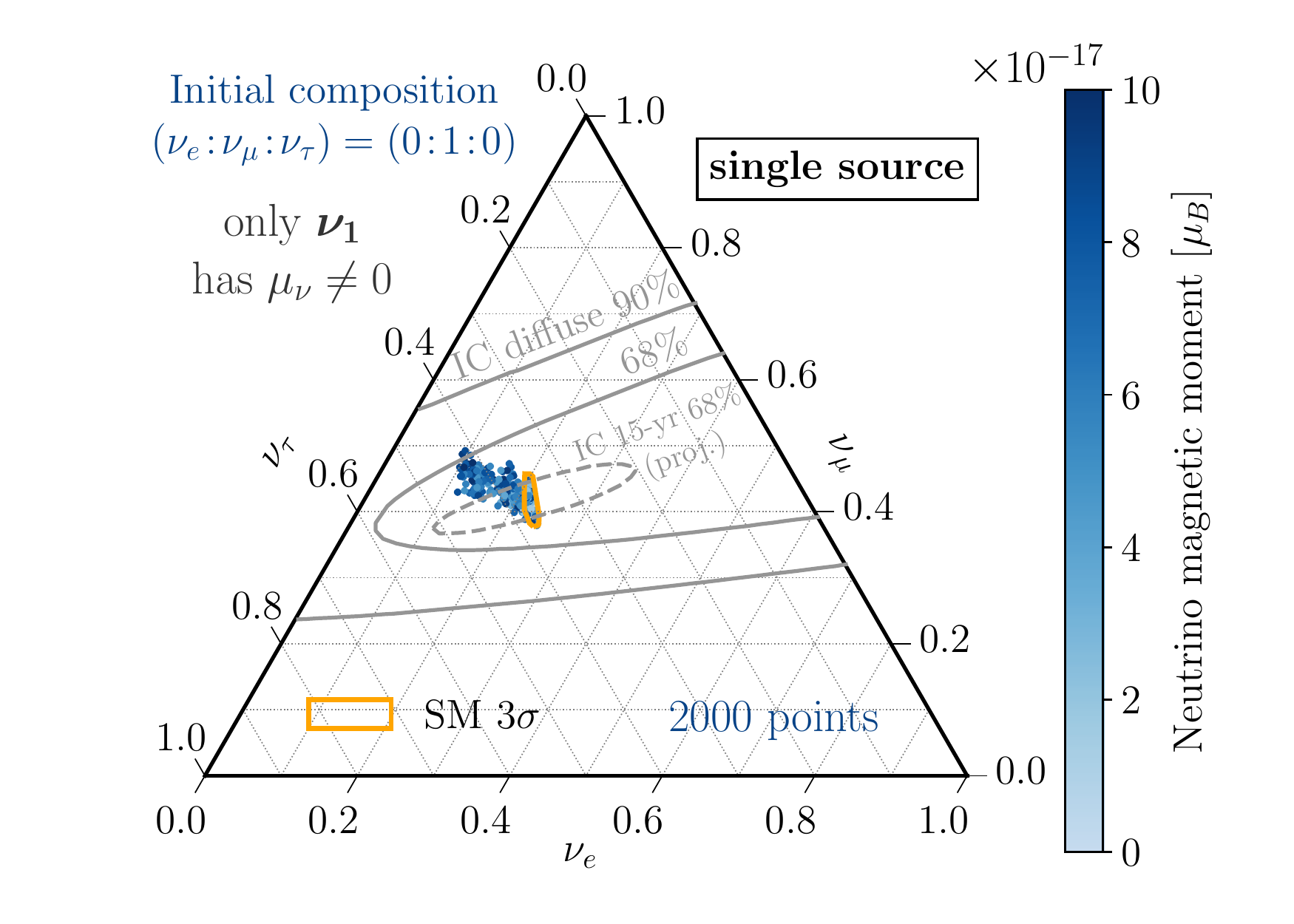} \\[-0.3cm]
        \includegraphics[width=0.9\columnwidth]{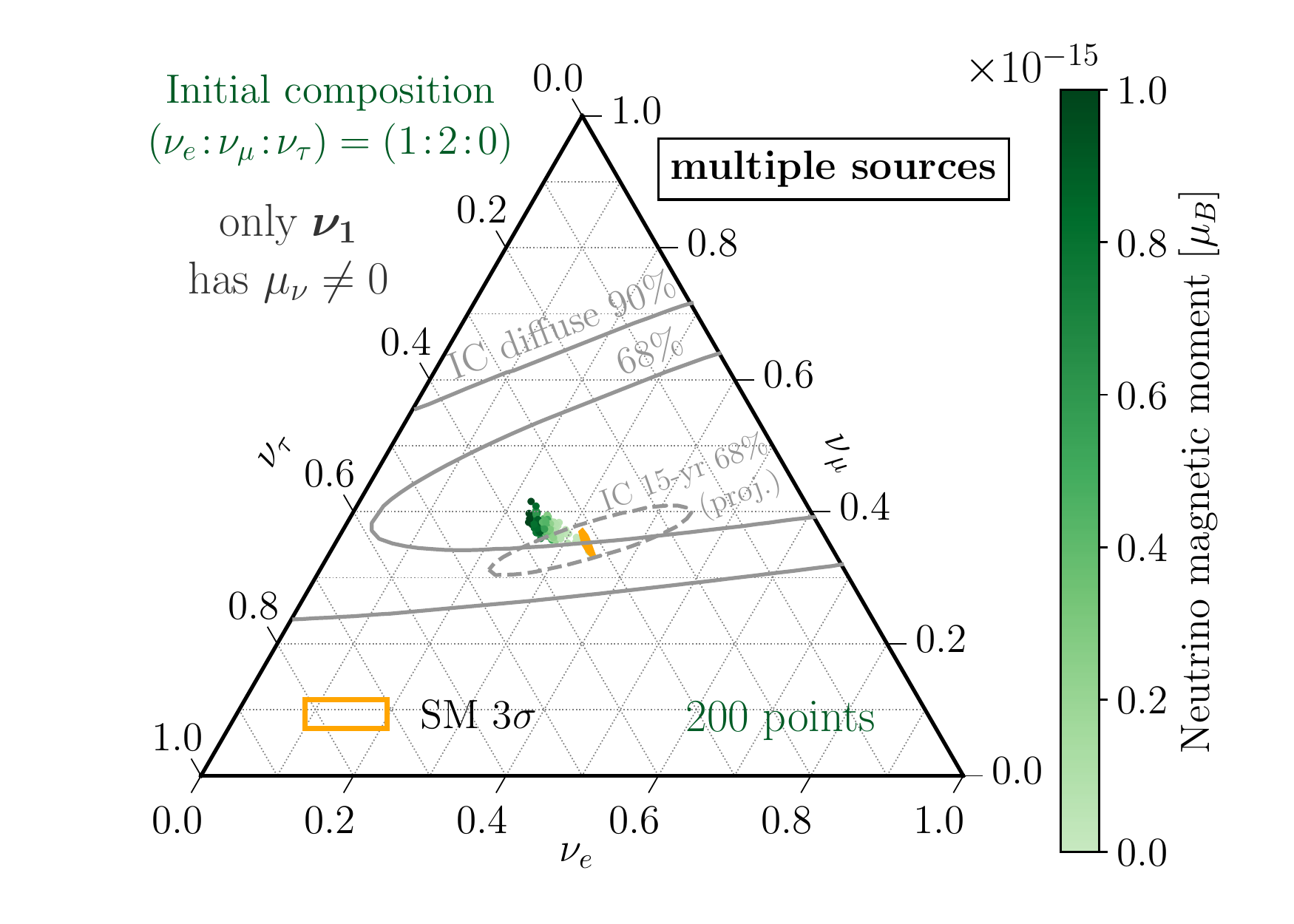} &
        \includegraphics[width=0.9\columnwidth]{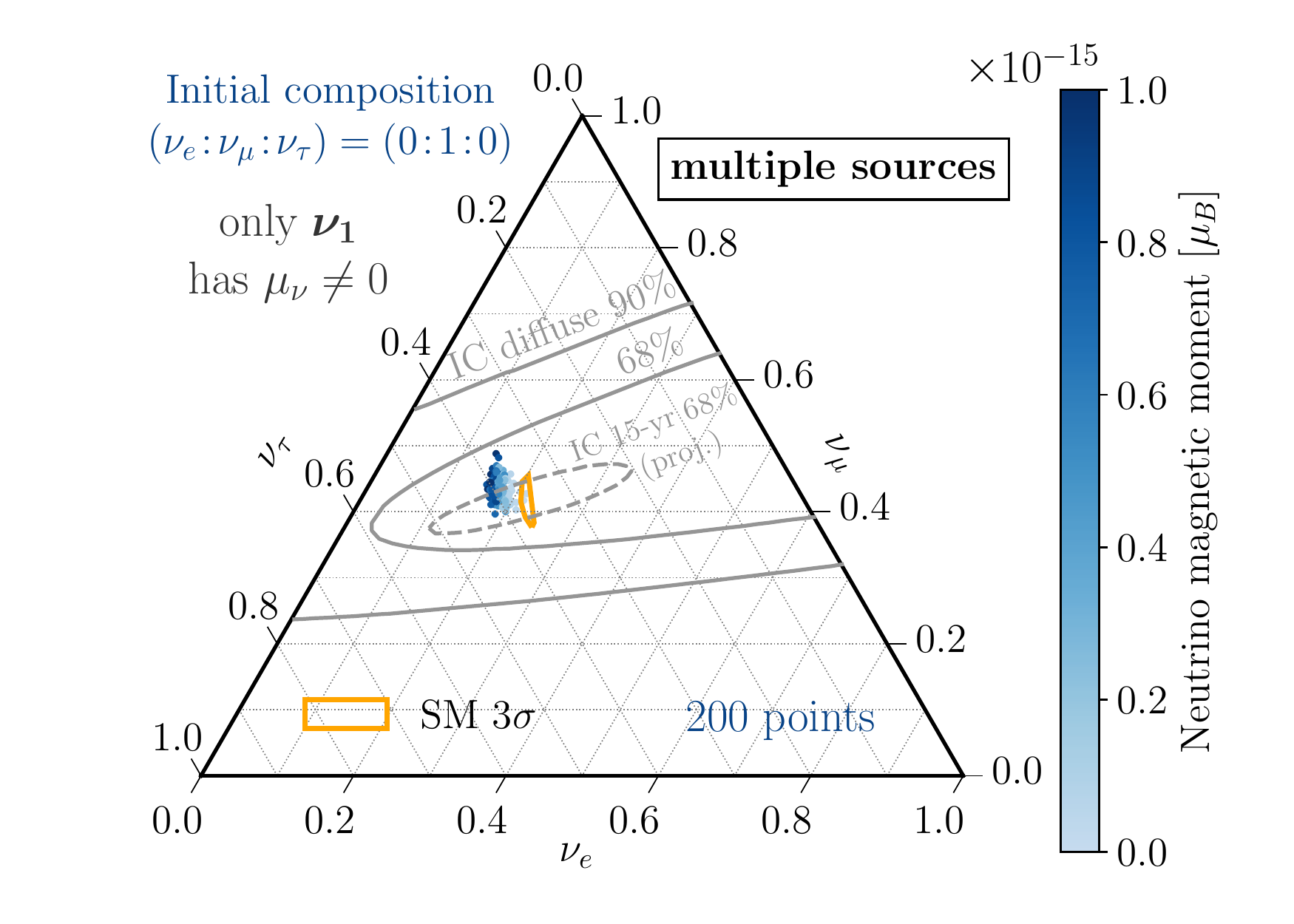}
    \end{tabular}
    \vspace{-0.5cm}
    \caption{Same as \cref{fig:flavour-triangles,fig:flavour-triangles-multi-source_mu_only}, but for a scenario where only the $\nu_1$ mass eigenstate carries non-zero magnetic moment.}
    \label{fig:flavour-triangles-nu1_only}
\end{figure*}

In \cref{fig:flavour-triangles-nu1_only}, we repeat our analysis for a different magnetic moment flavour structure, namely one where only the $\nu_1$ mass eigenstate carries non-zero $\mu_\nu$. Once again, deviations from the Standard Model flavour ratios are larger in the single-source case, but also for a diffuse flux, magnetic moments well below current limits can lead to deviations that should be observable in future IceCube analyses. This can be understood from the fact that the flavour composition of $\nu_1$ in the normal hierarchy case is about 50\% $\nu_e$. Removing $\nu_1$ from the neutrino flux via oscillations to $N_R$ thus disproportionately depletes the $\nu_e$ component at Earth, even after averaging over many magnetic field configurations.

Plots for other flavour structures of the magnetic moment matrix can be found on the GitHub repository accompanying this paper \cite{GitHubCode}.

\section{Magnetic Moment-Induced $\nu_L \leftrightarrow N_R$ Oscillations in Turbulent Magnetic Fields}
\label{sec:Posc-approx}

In the main part of this paper, we have evaluated the probability for $\nu_L \leftrightarrow N_R$ transitions in galactic and intergalactic magnetic fields by numerically solving the Schr\"odinger equation \eqref{eq:schroedinger}.  This is, however, rather time-consuming, especially for large $\mu_\nu$, where it requires tracking potentially many oscillation maxima and minima.  Also generating and storing the turbulent magnetic field configurations comes with a non-negligible overhead.

\begin{figure*}
    \centering
    \includegraphics[width=0.35\textwidth]{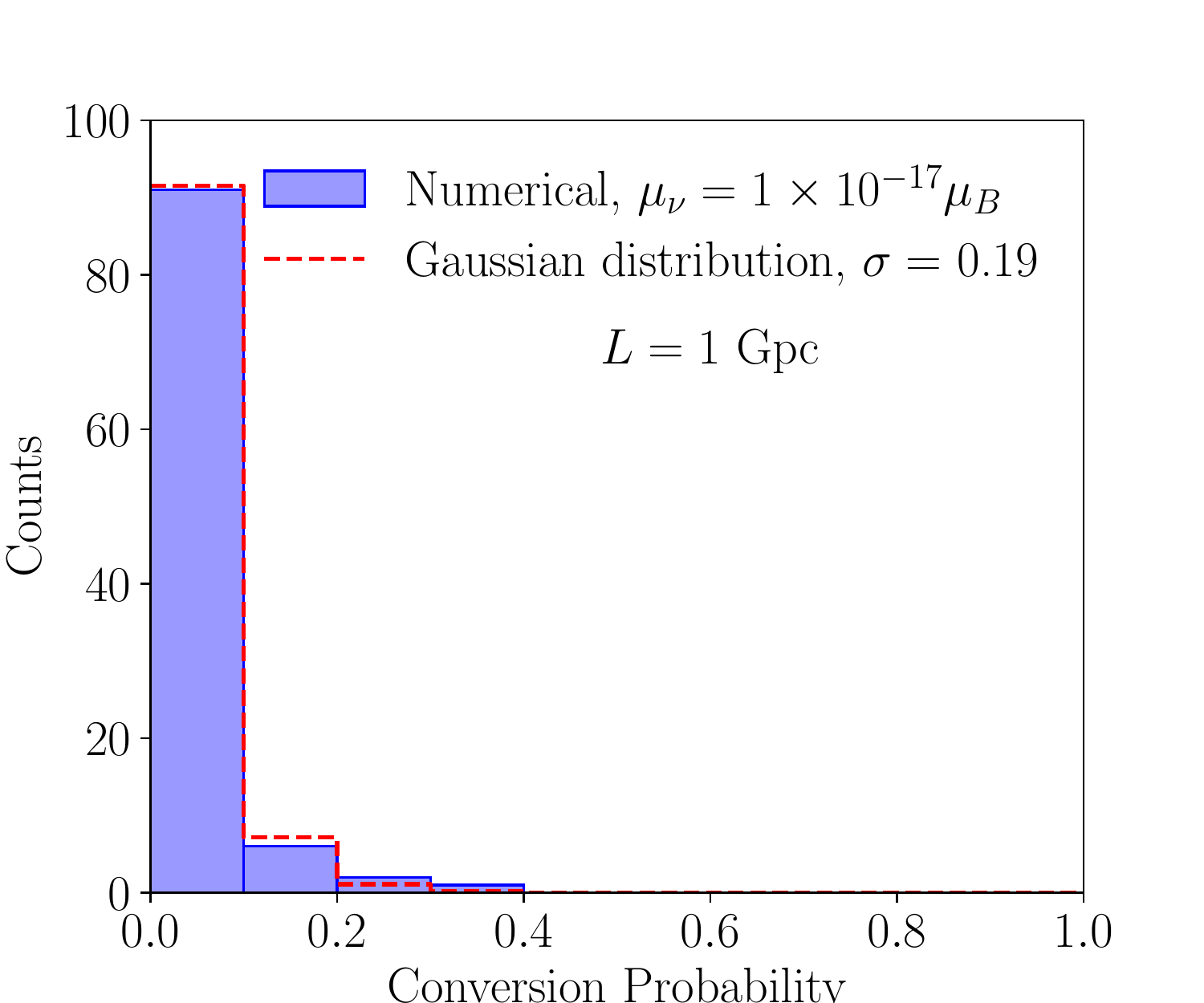}
    \hspace{-0.7cm}
    \includegraphics[width=0.35\textwidth]{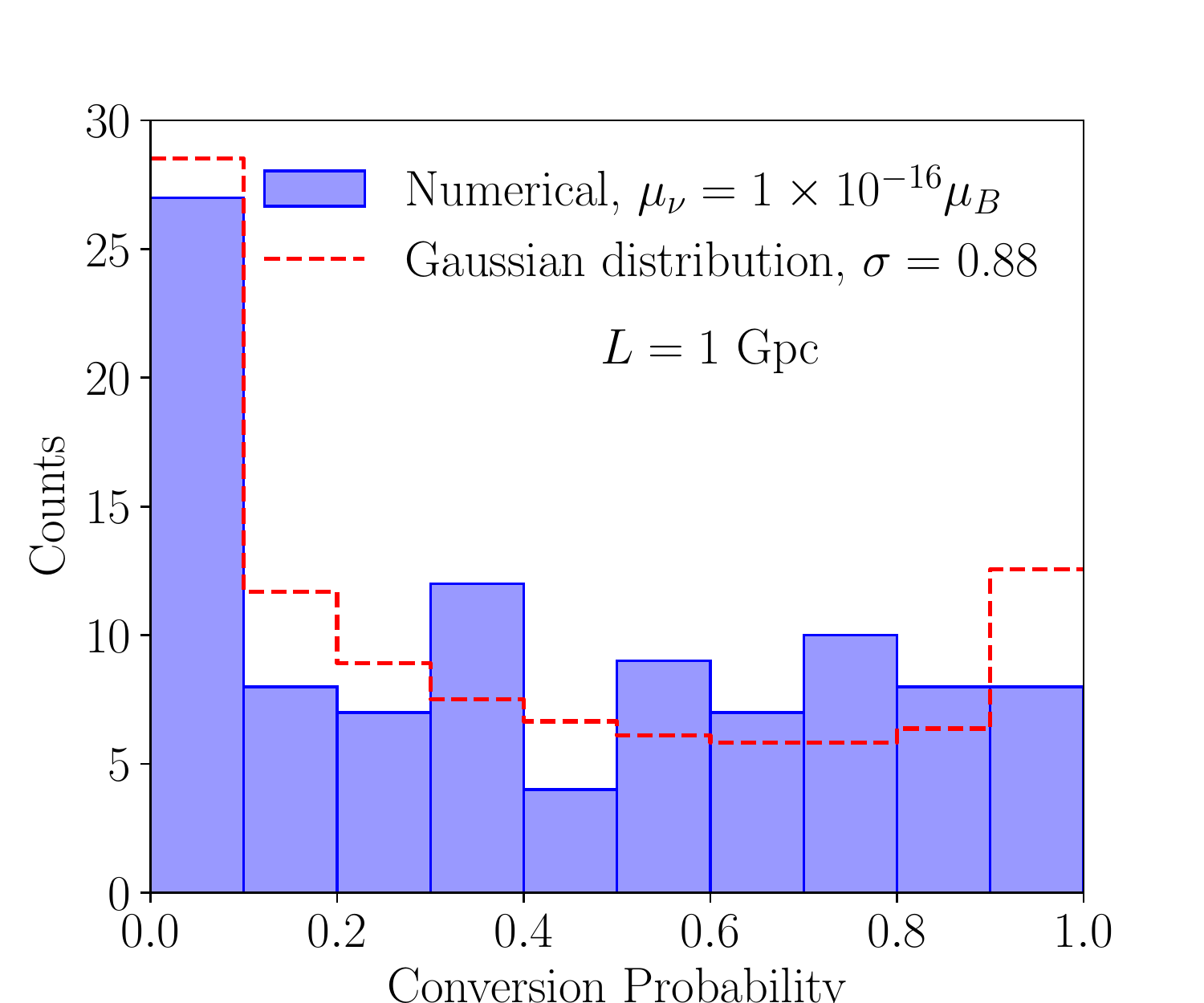}
    \hspace{-0.7cm}
    \includegraphics[width=0.35\textwidth]{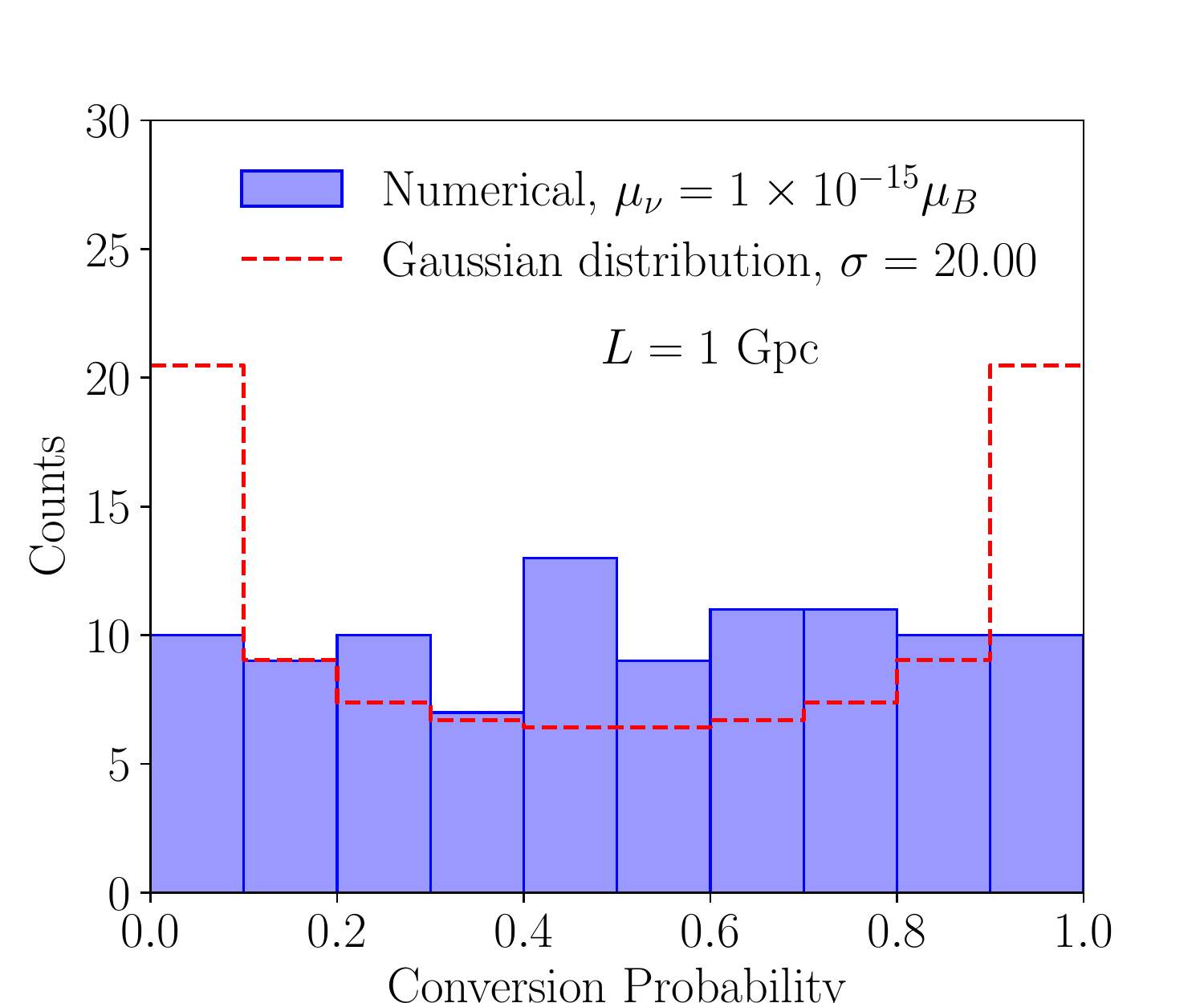}
    \caption{Comparison of the exact numerical results for the two-flavour $\nu_L \to N_R$ oscillation probability in a turbulent magnetic field to the approximate results discussed in \cref{sec:Posc-approx}. The blue-shaded histogram shows the distribution of numerically calculated $P_{\nu_L \to N_R}(L = \SI{1}{Gpc})$ values for 100 different extragalactic magnetic field profiles generated according to the procedure outlined in \cref{sec:B-field-extragalactic}. Unshaded red dashed histograms correspond to approximations based on drawing the oscillation phase $\varphi(L)$ from a Gaussian distribution with a width fitted to the numerical result. The three panels correspond to three different values of the neutrino magnetic moment, $\mu_\nu$.}
    \label{fig:Posc-approx}
\end{figure*}

For this reason, we have also investigated an approximate method for determining these probabilities, especially for the case of propagation through turbulent magnetic fields while using the two-flavour approximation. Taking into account the stochastic nature of these fields, the idea is to describe the oscillation probability as a $\sin^2$-function with a randomly chosen phase, $\varphi(t)$.  Specifically, we draw $\varphi(L)$ from a Gaussian distribution with a width that is fitted to the numerically calculated distribution. This is motivated by the observation that, for small $\mu_\nu$ (such that $\mu_\nu \ev{B_\perp} L \ll 1$), neutrinos will not have time to complete a single oscillation, even in the most favourable case that $\vec{B}_\perp(x)$ does not change direction along the line of sight. This extreme scenario, which maximises $\varphi(L)$, is statistically unlikely, while cancellations between regions with different field orientations are more likely. This behaviour is correctly modelled by the Gaussian distribution, as illustrated in the left and middle panels of \cref{fig:Posc-approx}.

In the opposite limit, $\mu_\nu \ev{B_\perp} L \gg 1$, the Gaussian approximation is less suitable.  Here, even small changes in $\vec{B}_\perp(t)$ lead to large changes in the oscillation phase. so one might speculate that the phase should be uniformly distributed. This would indeed be the case if the direction of $B_\perp$ did not change, so that the oscillation probability follows precisely a $sin^2$ law, as in \cref{eq:P-LtoR-varying-B-1}. For realistic field profiles, however, the behaviour of $P_{\nu_L \to N_R}$ is more involved, and in fact \cref{fig:Posc-approx} (right) shows that in this case a uniform distribution \emph{in the probability} provides the most accurate description.

\section{Detailed Derivation of the Decoherence Condition}
\label{sec:coherence-calculation}

        In \cref{sec:coherence}, we have heuristically argued that $\nu_L \to N_R$ transitions can occur coherently only if $m_N$ satisfies \cref{eq:coherence-cond}.  We will now corroborate the arguments given there by employing the wave packet formalism.  Our strategy is to compute the probability for $\nu_L \to N_R$ scattering on a background electron, assuming the initial and final state electron wave packets are identical. This corresponds to the regime of coherent conversion. For large $m_N$, the cross-section will be strongly suppressed, indicating the transition to the incoherent regime.  We are interested in the value of $m_N$ at which this transition happens.

We write the initial and final state  wave packets in the form
\begin{align}
	\ket{\psi} = \int \frac{d^3 p}{(2 \pi)^3 \sqrt{2 E_\vec{p}}} f (\vec{p}) \ket{\vec{p}},
\end{align}
where $\ket{\vec{p}}$ is a momentum eigenstate, and $E_\vec{p}$ is the corresponding energy.  The momentum eigenstates are normalised according to
\begin{align}
	\sprod{\vec{k}}{\vec{p}} = 2 E(\vec{k}) \,
	                           (2 \pi)^3 \delta^{(3)} (\vec{p} - \vec{k}) \,,
\end{align}
and the normalisation of the wave packet shape factor, $f(\vec{p})$ is
\begin{align}
    \int \frac{d^3 p}{(2 \pi)^3} \, |f(\vec{p})|^2 = 1 \,.
    \label{eq:wp-norm}
\end{align}
For the electrons, we choose a Gaussian wave packet with standard deviation $\sigma_p$ centred around $0$:
\begin{align}
	f_e (\vec{p}) = \bigg( \frac{2 \pi}{\sigma_p^2} \bigg)^{3/4}
	                \exp \bigg(\!\! -\frac{\vec{p}^{\, 2}}{4 \sigma_p^2} \bigg) \,.
\end{align}
We take the neutrino to be initially in a momentum eigenstate:
\begin{align}
	f_\nu (\vec{p}) = \frac{1}{\sqrt{V}} (2 \pi)^3
	                  \delta^{(3)} (\vec{p} - \vec{k}_\nu) \,,
	\label{eq:neutrino_wp}
\end{align}
where the prefactor is chosen to satisfy the normalisation condition, \cref{eq:wp-norm}.\footnote{The philosophy here is that we imagine for the moment that the process takes place in a box of finite volume $V$ and is restricted to a finite time interval, $T$.  $\delta$-functions should then be thought of as strongly peaked functions defined on these intervals, for instance $\delta(p^0) = \int_{-T/2}^{T/2} \! (dt/2\pi) \, e^{i p^0 t} = \sin(p^0 T/2) / (\pi p^0)$. To check that \cref{eq:neutrino_wp} indeed satisfies the normalisation condition, \cref{eq:wp-norm}, one can use the identities
\begin{align}
	\Big[(2 \pi)^3 \delta^{(3)} (\vec{p} - \vec{q}) \Big]^2
	    &= V (2 \pi)^3 \delta^{(3)} (\vec{p} - \vec{q}), \\
	\Big[(2 \pi) \delta (p^0 - q^0) \Big]^2
	    &= T (2 \pi) \delta (p^0 - q^0).
	\label{eq:delta_en_sq}
\end{align}}
For the neutrino in the final state, we use an analogous function.

The transition probability is then given by
\begin{widetext}
\begin{align}
	\mathcal{P} = \int \frac{d^3 p_N}{(2 \pi)^3} V \bigg|
	  \frac{1}{\sqrt{2 E_{\vec{k}_\nu} V}}
	  \frac{1}{\sqrt{2 E_{\vec{p}_N} V}}
	  \int \! \frac{d^3 k_e}{(2 \pi)^3} \frac{f_e(\vec{k}_e)}{\sqrt{2 E_{\vec{k}_e}}}         \frac{d^3 p_e}{(2 \pi)^3} \frac{f_e^*(\vec{p}_e)}{\sqrt{2 E_{\vec{p}_e}}} \mathcal{A}(k_\nu, k_e, p_N, p_e)
	\bigg|^2\,, 
\end{align}
where the subscripts $\nu$, $N$, $e$, refer to the active neutrino $\nu_L$, sterile neutrino $N_R$, and electron, respectively. We use the letter $k$ for initial state momenta and $p$ for final state momenta, and we already impose that the initial and final state electron wave packets should have the same shape, $f_e$.  The quantity $\mathcal{A}(k_\nu, k_e, p_N, p_e)$ is the scattering amplitude for momentum eigenstates, which we can obtain from the Feynman rules of the theory. As done in ref.~\cite{Beuthe:2001rc}, it is convenient to express $\mathcal{A}$ in coordinate space:
\begin{align}
	\mathcal{A}(k_\nu, k_e, p_N, p_e)
	  = \int \! d^4x \, \mathcal{M}_e(k_e, p_e) e^{-i (p_e - k_e) \cdot x}
	    \int \! d^4y \, \mathcal{M}_\nu(k_\nu, p_N) e^{-i (p_N - k_\nu) \cdot y} \, G(y-x)\,,
\end{align}
where $\mathcal{M}_e$ and $\mathcal{M}_\nu$ are the electron and the neutrino interaction amplitudes respectively, and the propagator, $G(y-x)$, is:
\begin{align}
	G(y-x) = \int\!\frac{d^4 p}{(2 \pi)^4} e^{-i p \cdot (y - x)}
	               \frac{i}{p^2 + i \epsilon}\,.
\end{align}
We first carry out the integrals over $\vec{k}_e$ and $\vec{p}_e$. We use the fact that $\mathcal{M}_e$ varies much more slowly as a function of $k_e$ and $p_e$ than both $f_e$ and the oscillating exponentials, hence it is justified to pull $\mathcal{M}_e$ out of the momentum integrals and replace it by its value at $k_e = p_e = 0$. Moreover, as the electron is non-relativistic, we can replace its energy by its mass. We thus arrive at two simple Gaussian integrals, each of which has the form
\begin{align}
	\int \! \frac{d^3p}{(2 \pi)^3}
	        \frac{f_{e}(\vec{p})}{\sqrt{2 E_{\vec{p}_e}}} e^{-i p \cdot x}
	  = \bigg( \frac{2 \sigma_p^2}{\pi} \bigg)^{3/4} \frac{1}{\sqrt{2 m_e}}
	     e^{-\sigma_p^2 \vec{x}^{\, 2} - i m_e t} \,.
\end{align}
Next is the integral over $x$:
\begin{align}
	\int \! d^4x \, e^{i p \cdot x} \bigg( \frac{2 \sigma_p^2}{\pi} \bigg)^{3/2}
	                \frac{1}{2 m_e} e^{- 2 \sigma_p^2 \vec{x}^{2}}
	    = 2 \pi \, \delta (p^0) \, \frac{1}{2 m_e} e^{-\frac{\vec{p}^2}{8 \sigma_p^2}} \,.
\end{align}
The integral over $y$ yields just a delta function,
\begin{align}
    \int \! d^4y \, e^{-i (p + p_N - k_\nu) y}
        = (2\pi)^4 \delta^{(4)}(p + p_N - k_\nu) \,,
\end{align}
which we can use to evaluate also the integral over $p$. For the transition probability, we are thus left with
\begin{align}
	\mathcal{P} =
	    \frac{1}{2 E_{\vec{k}_\nu}}
	    \frac{1}{(2 m_e)^2}
	    \mathcal{M}_e^2 \, \mathcal{M}_\nu^2
	    \bigg( \frac{2 \sigma_p^2}{\pi} \bigg)^3
	    \int \frac{d^3 p_N}{(2 \pi)^3} 
	    \frac{1}{2 E_{\vec{p}_N}}
	    e^{-\frac{(\vec{p}_N - \vec{k}_\nu)^2}{4 \sigma_p^2}} \,
	    \Big|
	        2 \pi \delta(p_N^0 - k_\nu^0)
	        \frac{1}{(\vec{p}_N - \vec{k}_\nu)^2}
    	\Big|^2 \,.
    \label{eq:P-decoh}
\end{align}
It would be straightforward to evaluate the remaining integral in spherical coordinates, using \cref{eq:delta_en_sq} to rewrite the squared delta function in terms of a single delta function. However, we can already read off the relevant physical conclusions from \cref{eq:P-decoh}. The delta function requires the $\nu_L$ and $N_R$ energies to be identical, and with this constraint, the exponential takes the form
\begin{align}
    e^{-\frac{(\vec{p}_N - \vec{k}_\nu)^2}{4 \sigma_p^2}}
        = \exp\bigg[ -\frac{2 E_\nu^2 - m_N^2 - 2 E_\nu \sqrt{E_\nu^2 - m_N^2} \cos\theta}{4\sigma_p^2} \bigg] \,.
\end{align}
\end{widetext}
Here, $E_\nu \equiv |\vec{k}_\nu| = \sqrt{|\vec{p}_N|^2 + m_N^2}$ is the common energy of $\nu_L$ and $N_R$, and $\theta$ is the angle between $\vec{k}_\nu$ and $\vec{p}_N$. The exponential is most strongly suppressed at $\cos\theta = -1$, and requiring no suppression corresponds to the condition that the exponent be larger than $\sim -1$. This translates into the condition
\begin{align}
    m_N < 2 \sqrt{\sigma_p (E_\nu - \sigma_p)} \,,
    \label{eq:coherence-cond-2}
\end{align}
which is precisely \cref{eq:coherence-cond}.

\bibliographystyle{JHEP}
\bibliography{refs}

\end{document}